\documentclass[reprint,amsmath,amssymb,aps,pra,superscriptaddress,raggedbottom]{revtex4-2}
\usepackage{graphicx}
\usepackage{dcolumn}
\usepackage{bm}
\usepackage{xcolor}
\usepackage{hyperref}
\usepackage{enumitem}
\usepackage[none]{hyphenat}
\DeclareMathAlphabet\mathbfcal{OMS}{cmsy}{b}{n}
\usepackage{multirow}
\usepackage{booktabs}

\begin{document}
	
	\title{Formation of axially modulated plasma strings\\ by filamentation of interfering femtosecond Bessel beams}
	
	\author{Fatemeh Mansourimanesh}
	\affiliation{
		Department of Electrical Engineering, University of Tehran, Tehran, Iran
	}
	
	\author{Amirreza Sadeghpour}

	\author{Daryoush Abdollahpour}
	\email{dabdollahpour@iasbs.ac.ir}
	\affiliation{
		Department of Physics, Institute for Advanced Studies in Basic Sciences (IASBS), Zanjan 45137‑66731, Iran}
	\date{\today}
	
\begin{abstract}
\noindent We numerically investigate the formation of axially modulated plasma strings through the filamentation of two interfering femtosecond Bessel beams. The constituent Bessel beams have different central spot sizes and propagate collinearly. Our results show that filamentation of these beams in air leads to the formation of intense axially modulated light filaments, and corrugated plasma strings with high modulation depths and tunable modulation periods. We find that the modulation periods of the intensity and the plasma density are identical and comparable to the modulation period of the intensity in the linear propagation regime. Furthermore, we show that, in the nonlinear propagation regime, the modulation period is independent of the pulse energy and can be tuned by selecting appropriate central spot sizes for the interfering Bessel beams. Finally, we propose a simple interferometric arrangement using a single axicon to generate two interfering Bessel beams with adjustable spot sizes enabling the creation of axially-modulated plasma strings with tunable periodicity at high input powers. 
\end{abstract}
	
	\maketitle
	
\section{\label{sec1} Introduction}
	Femtosecond filaments are the outcome of a dynamic equilibrium involving several linear and nonlinear effects including self-focusing, plasma defocusing, diffraction, dispersion, and losses due to multiphoton absorption and absorption by plasma \cite{CouaironPhysRep07}. Filamentation has several interesting features, including intensity clamping (at $\sim$$10^{13}$ W/cm$^{2}$) over an almost constant diameter persisting over remarkably long distances beyond several Rayleigh ranges, generation of underdense plasma channels, supercontinuum generation, and conical emission \cite{CouaironPhysRep07, TzortzakisPRL01}. Owing to these features, light filaments have been utilized in several applications such as laser-guided lightning \cite{HouardNP23}, remote sensing \cite{KasparianSci03}, few-cycle pulse generation \cite{CouaironOL05, MitroOptica16}, THz generation \cite{XiePRL06, KimNP08}, high-harmonic generation \cite{SteinNJP11} and efficient third-harmonic generation \cite{SuntsovOE09, SuntsovPRA10}. Moreover, it has been shown that two intersecting light filaments can form a plasma grating \cite{SuntsovAPL09,ShiPRL11} that can be used for several applications of ultraintense laser beams \cite{LehmannPRE19, EdwardsPRAp22, EdwardsPRL24}. 
	
Previous studies indicate that axially modulated laser-generated plasma channels can be utilized as slow-wave structures for direct laser acceleration of charged particles or for generating a broad spectrum of electromagnetic radiation \cite{LayerPRL07}. In an axially uniform plasma waveguide, the phase velocity mismatch between the driving laser pulse and the secondary electromagnetic radiation (or the particles) restricts an efficient interaction of the waves. However, in a corrugated plasma waveguide, the laser field is slowed down and an efficient phase matching between the interacting waves is achieved. The phase matching between the interacting waves in an axially modulated plasma channel can equivalently be described by quasi-phase-matching (QPM) \cite{LayerOE09, YoonPRL14}. The corrugated plasma channels have been proposed or used for plasma accelerator injectors \cite{KatTNS85, DahiyaAPL10}, THz generation \cite{AntonsenPP07, PearsonPRE11}, and relativistic acceleration of electrons \cite{YorkPRL08, GuptaPRE11, YoonPRL14}. Furthermore, an axially corrugated plasma structure can be used for increasing the efficiency of third-harmonic generation in plasmas by providing the possibility of quasi-phase-matching between the fundamental wave and its third-harmonic \cite{SuntsovPRA10}.
	
	Several approaches have been employed for generating axially modulated plasma channels. For instance, in Ref. \cite{LayerOE09} an intense picosecond pulse is focused by an axicon into a density-modulated cluster jet (produced via wire obstruction) to create a preformed corrugated plasma waveguide. In Ref. \cite{LinPP06}, the amplitude of a high-energy (multi-millijoule) femtosecond laser pulse is modulated by a one-dimensional liquid-crystal spatial light modulator (LC-SLM), and the demagnified image of the modulated intense beam is employed to generate a preformed high-modulation-depth plasma channel for another traversing beam propagating through the modulated plasma structure. However, such high-power pulses are likely to damage LC-SLMs \cite{HineOL16}. In contrast, in Ref. \cite{HineOL16} a low-energy pulse is sent to an SLM for phase-front modulation. The modulated low-energy laser pulse is then interferometrically combined with a high-energy pulse to sculpt the intensity pattern of the superposed fields, which is subsequently focused by an axicon to produce axially corrugated plasma density. However, the significant amplitudes difference between the constituent beams leads to a reduced modulation depth in the axial intensity and consequently in the plasma density. 
	
Moreover, it has been shown that the superposition of two Bessel beams in the linear propagation regime leads to a self-imaging effect in the superimposed field \cite{CerdaOL98, CerdaOE98}. Regarding the fact that nonlinear propagation dynamics of optical beams are strongly influenced by their specific energy flux in the linear propagation regime \cite{AbdollahpourOE09}, it could be anticipated that a self-imaging effect would also occur during the filamentation of the superposition of ultrashort-pulse Bessel beams, potentially leading to the formation of an axially modulated plasma structure. 

Furthermore, the filamentation of femtosecond Bessel beams in various media has been the subject of extensive research \cite{GaizOL06, RoskeyOE07, PolesanaPRA08, AbdollahpourOE09, AkturkOptComm09, PolynkinOE09, DotaPRA12, CouaironSPIE13, PorrasPRA15}. It has been demonstrated that several parameters---such as the cone angle of the conical wavefront of the Bessel beam (or equivalently that of the axicon), width of the incident beam on the axicon, initial power of the beam, and the strength of nonlinear losses---determine the nonlinear propagation regime of a Bessel beam in transparent media. In particular, for 800 nm femtosecond Bessel beams propagating in air, it is well established that small cone angles (corresponding to larger spot sizes of the Bessel beam), and high input powers lead to so-called \emph{unsteady filamentation}, characterized by irregular modulations of intensity and electron density along the propagation direction \cite{PolesanaPRA08, PorrasPRA15}. In this regime, the peak intensity and electron density reach values comparable to those associated with the filamentation of conventional Gaussian beams. Conversely, for large cone angles (i.e., smaller central spot sizes of the Bessel beam), \emph{steady filamentation} occurs, with nearly uniform intensity and electron density maintained along the propagation direction \cite{PolesanaPRA08, PorrasPRA15}. Notably, Ref. \cite{PolesanaPRA08} demonstrates that over a wide range of parameters, it is possible within the steady filamentation regime to generate Bessel filaments with various peak intensities and electron densities. Interestingly, it is shown that for large cone angles,  light filaments can be formed with negligible plasma densities (well below $10^{17}$ cm$^{-3}$), and hence with minimal pulse reshaping due to plasma defocusing. 
	
Through numerical investigations, here we demonstrate the possibility of generating longitudinally modulated plasma structures with significant modulation depths by filamentation of two superposed femtosecond Bessel beams in air. It is also shown that the attributes of the modulated plasma structures  (such as the modulation period and depth) can be tuned by adjusting the parameters of the interfering Bessel beams. Moreover, we propose a simple interferometric arrangement with a single axicon to achieve a high-modulation-depth plasma structure without a need for an SLM.  

\section{\label{sec2} Model and Methods}
The electric field envelope of the beam, at a propagation distance $z=0$, is modeled as a superposition of two Bessel beams with an initial phase difference of $\pi$, and with a Gaussian spatiotemporal envelope 
\begin{equation}
	\begin{split}
	\mathcal{E}(\rho, t, z=0)=&\mathcal{E}_0 \left[J_0(\rho/s_1) -J_0(\rho/s_2)\right]\\
	&\times\exp\left\{-\left[\left(\frac{\rho}{w}\right)^{2}+\left(\frac{t}{t_{0}}\right)^{2}\right]\right\},
	\label{eq1}
	\end{split}
\end{equation}
where $\mathcal{E}_0$ is the amplitude, $J_{0}(\cdot)$ is the zeroth-order Bessel function of the first kind, $\rho$ is the radius in the transverse plane, $s_1$ and $s_2$ are scale factors proportional to the widths of the central spots of the interfering Bessel beams \cite{Durnin87}, $w$ is the width of the spatial Gaussian envelope and the beam is assumed to have a Gaussian temporal profile with a pulse width of $t_0$. The transverse wavenumber of each Bessel beam is given by $k_{\bot}~=~k\sin(\theta)=~1/s$, where $\theta$ is the cone angle of the Bessel beam. The longitudinal wavenumber is therefore given by $ k_z ~=~\sqrt{k^{2}-\smash{k_{\bot}^{2}}} = k\cos(\theta)$, with $k=2\pi/\lambda$ representing the total wavenumber of the beam. It has been shown that the intensity of the superposition of two collinearly-propagating zeroth-order Bessel beams with equal amplitudes, in linear propagation regime, is given by \cite{CerdaOL98, CerdaOE98, McGloinOL03}
\begin{equation}
\begin{split} 
		I(\rho,z) = &J_0^2(k_{\bot_1}\rho)+J_0^2(k_{\bot_2}\rho)\\
		&+2J_0(k_{\bot_1}\rho)J_0(k_{\bot_2}\rho)\cos\left[\left(k_{z_1}-k_{z_2}\right)z+\gamma\right]
		\label{eq1p}
	\end{split}
	\end{equation}
	where $z$ is the propagation distance, and $\gamma$ is the initial phase difference between the two beams. The modulation period of the axial intensity modulation is therefore found as 
	\begin{equation}
	\Lambda = \frac{2\pi}{|k_{z_1}-k_{z_2}|}\simeq \frac{4\pi ks_{1}^{2}s_{2}^{2}}{|s_{1}^{2}-s_{2}^{2}|}	
	\label{eq1pp}
	\end{equation} 

\begin{figure*}[ht!]
	\centering
	\includegraphics[width=2\columnwidth]{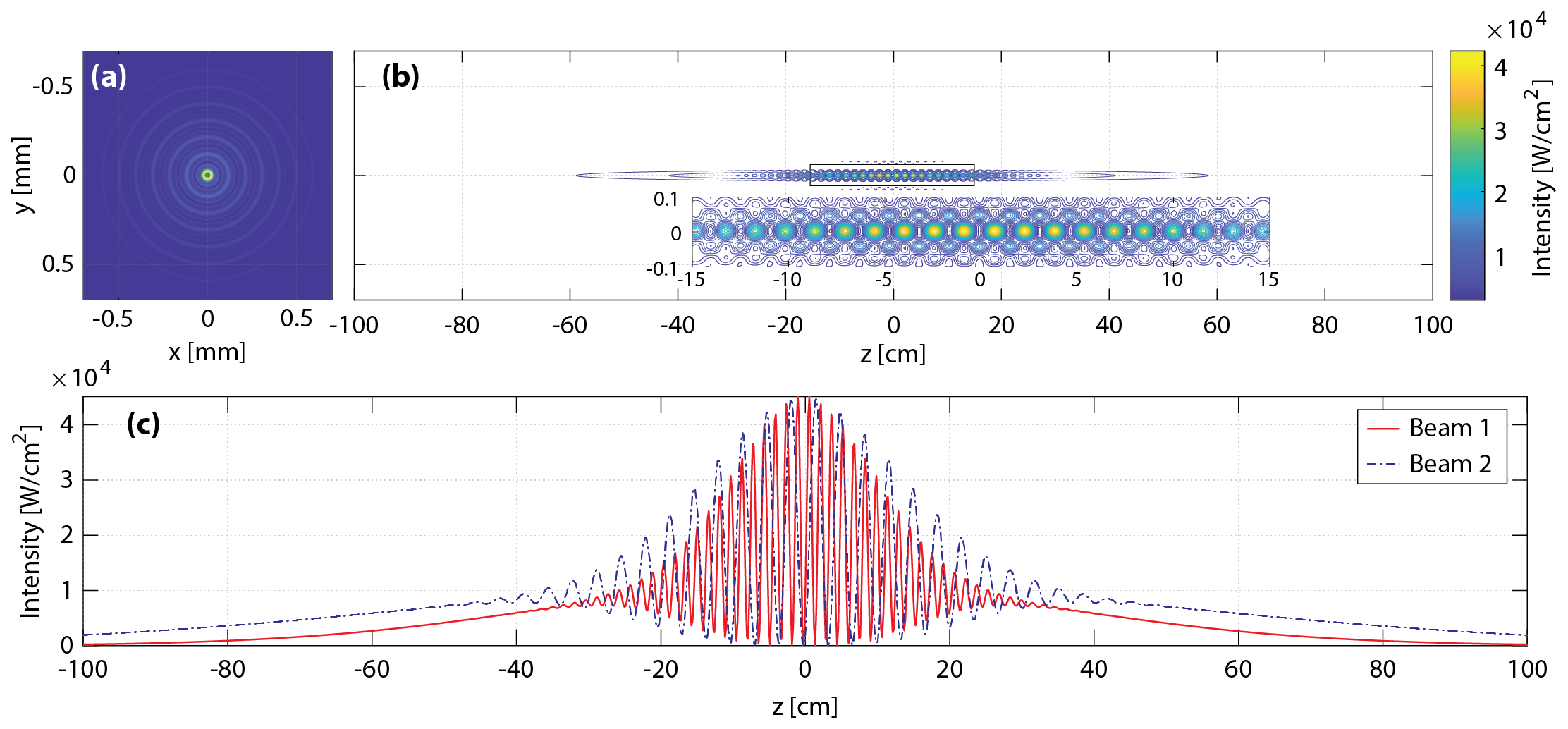}
	\caption{(a) Transverse intensity profile of \emph{Beam 1} at $z=0$; (b) Longitudinal intensity profile of \emph{Beam 1} in the linear propagation from $z=-1\ \mathrm{m}$ to $z=1\ \mathrm{m}$;(c) Axial intensity profiles of \emph{Beam 1} (solid curve), and \emph{Beam~2} (dashed curve). Parameters of the beams are as the following: \emph{Beam 1}: $s_{1}=60\ \mu\mathrm{m}$, $s_2=12\ \mu \mathrm{m}$, and \emph{Beam~2}: $s_1=90\  \mu \mathrm{m}$, $s_2=18\ \mu\mathrm{m}$; both beams have the pulse energy of 1 nJ, and the same spatiotemporal envelope with $w=1.5\ \mathrm{mm}$, and $t_0=50\ \mathrm{fs}$.}
	\label{fig1}
\end{figure*}

The transverse intensity profile of the superposed beam at $z=0$, with a wavelength of $\lambda=800~\text{nm}$, $w=1.5~\text{mm}$, $t_0=50$~fs, $s_1=60~\mu\text{m}$, $s_2=12~\mu\text{m}$, that will be referred to as \emph{Beam~1} hereafter, is shown in Fig.~\ref{fig1}(a). To investigate the linear and nonlinear propagation of the superimposed beam, the field envelope--given in Eq.~\eqref{eq1}-- was back propagated for 1~m (i.e., from $z=0$ to $z=-1$~m), and then the field envelope at $z=-1$~m was propagated forward over a propagation length of 2~m. The intensity contours of the linear propagation of \emph{Beam~1} in air, over a propagation range of $z=-1$~m to $z=1$~m is presented in Fig.~\ref{fig1}(b), and its axial profile is depicted in Fig.~\ref{fig1}(c) (solid curve). The high-contrast modulation of the intensity of the superimposed beam around $z=0$ is attributed to successive constructive and destructive interference between the two constituent Bessel beams during the propagation, as described by Eq.~\eqref{eq1p}.

For comparison, the axial intensity profile of the linear propagation of a superposition of two Bessel beams with different widths of $s_1=90~ \mu\text{m}$ and $s_2=18~\mu\text{m}$, but identical other parameters, are also depicted in Fig.~\ref{fig1}(c) (dashed curve). The latter beam will be referred to as \emph{Beam~2}, hereafter. For the linear propagation, the pulse energies of both beams were set to 1~nJ. A comparison of the two graphs indicates that the periodicity of the axial modulation can be tuned by adjusting the central spot sizes of the interfering components (i.e., sub-beams), $s_1$ and $s_2$, as anticipated from Eq.~\eqref{eq1pp}. From the numerical propagation profiles, the modulation periods, $\Lambda$, of \emph{Beam~1} and \emph{Beam~2} were measured to be 15~mm and 33.5~mm, respectively. These measurements are consistent with Eq.~\eqref{eq1pp} with a tolerance of $\sim$1\%.
The modulation depth of the intensity in the linear regime can be controlled by adjusting the relative amplitudes of the two interfering Bessel components and maximum modulation depth is achieved when the two components have equal initial amplitudes, as assumed in Eq.~\eqref{eq1}. Moreover, the phase difference between the interfering components (i.e., $\gamma$) determines the phase of the axial modulation and it can be arbitrarily tuned for specific applications.

Since the nonlinear propagation of optical beams is closely linked to their linear power flux \cite{AbdollahpourOE09, AbdollahpourPRL10, Vuong2006, Dergachev2014}, peculiar nonlinear propagation dynamics were expected for the filamentation of interfering femtosecond Bessel beams. For the numerical investigation of the nonlinear propagation of the beams, we used the classical model for the filamentation of ultrashort laser pulses. According to this model, the spatiotemporal evolution of the electric field envelope of the beam, in a time frame moving with the pulse along the propagation, is mathematically described as \cite{CouaironPhysRep07}
\begin{equation}
	\begin{split}
		\frac{\partial \mathcal{E}}{\partial z}=&\frac{i}{2k}\nabla_\bot^2\mathcal{E}-i\frac{k^{\prime \prime}}{2}\frac{\partial^2 \mathcal{E}}{\partial t^2}+i k n_2 |\mathcal{E}|^2 \mathcal{E}\\&-\frac{\sigma}{2}\left(1+i \omega_0 \tau_c\right) \rho_e \mathcal{E}-\frac{\beta_K}{2}|\mathcal{E}|^{(2K-2)} \mathcal{E},
	\label{eq2}
	\end{split}
\end{equation}

\noindent where $\omega_{0}$ is the carrier frequency of the field, $k=n_{0}\omega_{0}/c$ is the wavenumber, $n_{0}$ is the refractive index of the propagation medium at $\omega_{0}$, $c$ is the speed of light in vacuum, $\nabla_\bot^2$ denotes the transverse Laplacian operator, $k^{\prime \prime}$ is the group velocity dispersion (GVD) coefficient, $n_{2}$ is the coefficient of the nonlinear Kerr index, $\sigma$ is the plasma absorption cross-section, $\tau_c$ is the electron collision time, $\rho_{e}$ is the number density of electrons, $\beta_{K}$ is the multiphoton ionization coefficient, and $K$ is the number of photons involved in the multiphoton ionization process. The terms on the right-hand side of Eq.~\eqref{eq2} represent diffraction, group velocity dispersion, instantaneous Kerr self-focusing, plasma absorption and defocusing, and multiphoton absorption, respectively. Equation~\eqref{eq2} is coupled with the rate equation of electron density \cite{CouaironPhysRep07, CouaironPRA2003}
\begin{equation}
	\frac{\partial \rho_{e}}{\partial t}=W(I)(\rho_{\text{nt}}-\rho_{e}),
	\label{eq3}
\end{equation}

\noindent where $\rho_{\text{nt}}$ is the neutral molecule density of oxygen in  atmospheric pressure air, and $W(I)=\sigma_KI^K$ is the multiphoton ionization rate, with $I$ representing the intensity, and $\sigma_K$ denoting the multiphoton absorption cross-section. The rate equation is an ordinary differential equation that is solved via trapezoidal integration at each numerical step along the propagation direction. The calculated electron density at each step is then plugged in Eq.~\eqref{eq2} to account for the effect of the plasma. 

\begin{table}
		\caption{Physical parameters of the numerical simulations.}
		\label{tab1}
		\begin{ruledtabular}
		\begin{tabular}{ll}
		\textrm{Quantity}& \textrm{Value and unit}\\
		\colrule
		\\[-7pt]
			$K$ & $8$ (for oxygen) \\
			$\sigma_K$ & $3.7\times10^{-96}$[s$^{-1}$cm$^{16}$W$^{-8}]$ \\
			$n_2$ & $4 \times 10^{-19}$ [cm$^2 $W$^{-1}]$\\
			$\beta_K$ & $1.8\times10^{-94}$ [cm$^{13}$W$^{-7}]$ \\
			$\tau_c$ & 350 [fs] \\
			$k^{\prime\prime}$ & $0.2 $ [fs$^2$ cm$^{-1}$]  \\
			$\sigma$ & $5.47\times 10^{-20}$ [cm$^2$]  \\
			$\rho_\text{nt}$ & $5\times 10^{18}$ [cm$^{-3}$] (for O$_2$)  
			
		\end{tabular}
		\end{ruledtabular}
\end{table}

Equation~\eqref{eq2} is numerically solved using a spectral extended Crank-Nicolson scheme \cite{CouaironEPJ2011}. In this approach, the spatial finite-difference Crank-Nicolson scheme is applied for each frequency component of the field envelope, and the nonlinear terms are accounted for by a second-order Adams-Bashforth approximation. The physical space of simulation in the transverse directions is $60~\text{mm}$ with a uniform grid spacing of $1.6~\mu\text{m}$. In the time domain, we implemented a 1~ps-wide time window with a step size of 0.5~fs. Moreover, the step size along the direction of propagation, $z$, is set to $32~\mu\text{m}$. 

\section{\label{sec3} Results and Discussions}
 Filamentation of the superposed Bessel beams, \emph{Beam~1}, \emph{Beam~2}, and a third beam, \emph{Beam~3} (with $s_1 = 75~ \mu\text{m}$ and $s_2 = 15~ \mu\text{m}$, and hence $\Lambda = 23.1$~mm) in air, were numerically simulated with the physical parameters specified in Table \ref{tab1}. Furthermore, we conducted several simulations to investigate the power dependence of the nonlinear propagation of the beams with initial pulse energies of $1.03~\text{mJ}$ for all three beams; and a pulse energy of  $1.53~\text{mJ}$ for \emph{Beam~2}. The former pulse energy corresponds to a peak power of $\sim$6.45$P_\text{cr}$ for the three beams, while the latter pulse energy corresponds to $P=9.65~P_\text{cr}$ for \emph{Beam~2}; where $P_\text{cr}\simeq2.5~\text{GW}$ is the critical power of self-focusing in air at $\lambda=800~\text{nm}$ with $n_2=4\times 10^{-19}~\text{cm}^2\text{W}^{-1}$ \cite{CouaironPRA2003}.
 
 \begin{figure}[t!]
	\centering
	\includegraphics[width=1\columnwidth]{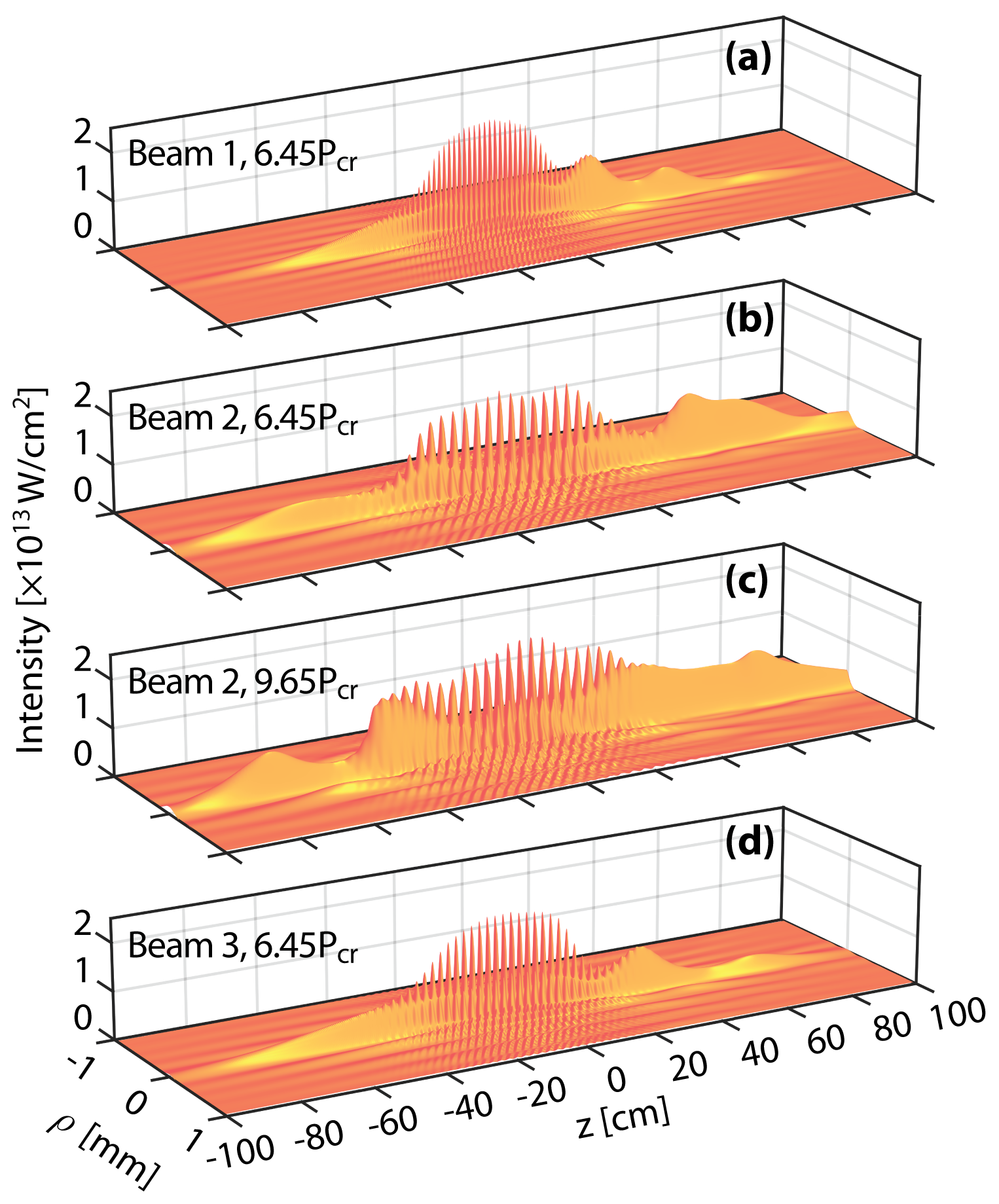}
	\caption{Evolution of the time-averaged intensity during the nonlinear propagation for (a) \emph{Beam~1} with an input power of $6.45\times P_{\text{cr}}$, (b) \emph{Beam~2} with an input power of $6.45\times P_{\text{cr}}$, (c) \emph{Beam~2} with an input power of $9.65\times P_{\text{cr}}$, and (d) \emph{Beam~3} with an input power of $6.45\times P_{\text{cr}}$. Here, $\rho$ denotes the radial distance from the propagation axis, $z$.}
	\label{fig2}
\end{figure} 
 
The evolution of the time-averaged intensity along the nonlinear filamentary propagation of the three beams, with the two pulse energies is shown in Fig.~\ref{fig2}. The modulation of the intensity along the direction of propagation is seen; indicating that the modulation is inherited from the linear propagation characteristics of the beams. The peak intensity along the propagation is clamped to $\sim$2$\times 10^{13}$~W/cm$^2$, consistent with the peak clamped intensity of the filamentation of femtosecond pulses \cite{CouaironPhysRep07, ZouPRA23}, and also with the peak intensity and corresponding peak fluence in filamentation of femtosecond Bessel beams in air \cite{RoskeyOE07, PolesanaPRA08}. Additionally, comparing the nonlinear propagation of \emph{Beam~1} (Fig.~\ref{fig2}(a)), \emph{Beam~2} (Fig.~\ref{fig2}(b-c)), and \emph{Beam~3} (Fig.~\ref{fig2}(d)) it is easily verified that the intensity modulation period in the nonlinear regime can also be adjusted by varying the sub-beam parameters, $s_1$ and $s_2$.
	
Moreover, a comparison of the nonlinear propagation of \emph{Beam~2} at two different pulse energies (Fig.~\ref{fig2}(b) and Fig.~\ref{fig2}(c)), reveals that increasing the input power does not affect the modulation period, but extends the filamentation length. \emph{Beam~2}, composed of Bessel beams with larger central spots (i.e., smaller cone angles), exhibits additional intensity peaks beyond the modulated region. This arises because larger central spots undergo local Kerr self-focusing, reshaping and creating high-intensity regions \cite{PolesanaPRA08}. These effects become more pronounced at higher input powers (Fig.~\ref{fig2}(c)).

To explore the potential link between the filamentation of interfering Bessel beams and that of their individual constituent sub-beams, we also investigated the nonlinear propagation dynamics of the individual Bessel sub-beams forming each of the three composite beams. The results of these investigations along with a detailed description are provided in Appendix~\ref{AppndxA}.  

\begin{figure}[t!]
	\centering
	\includegraphics[width=1\columnwidth]{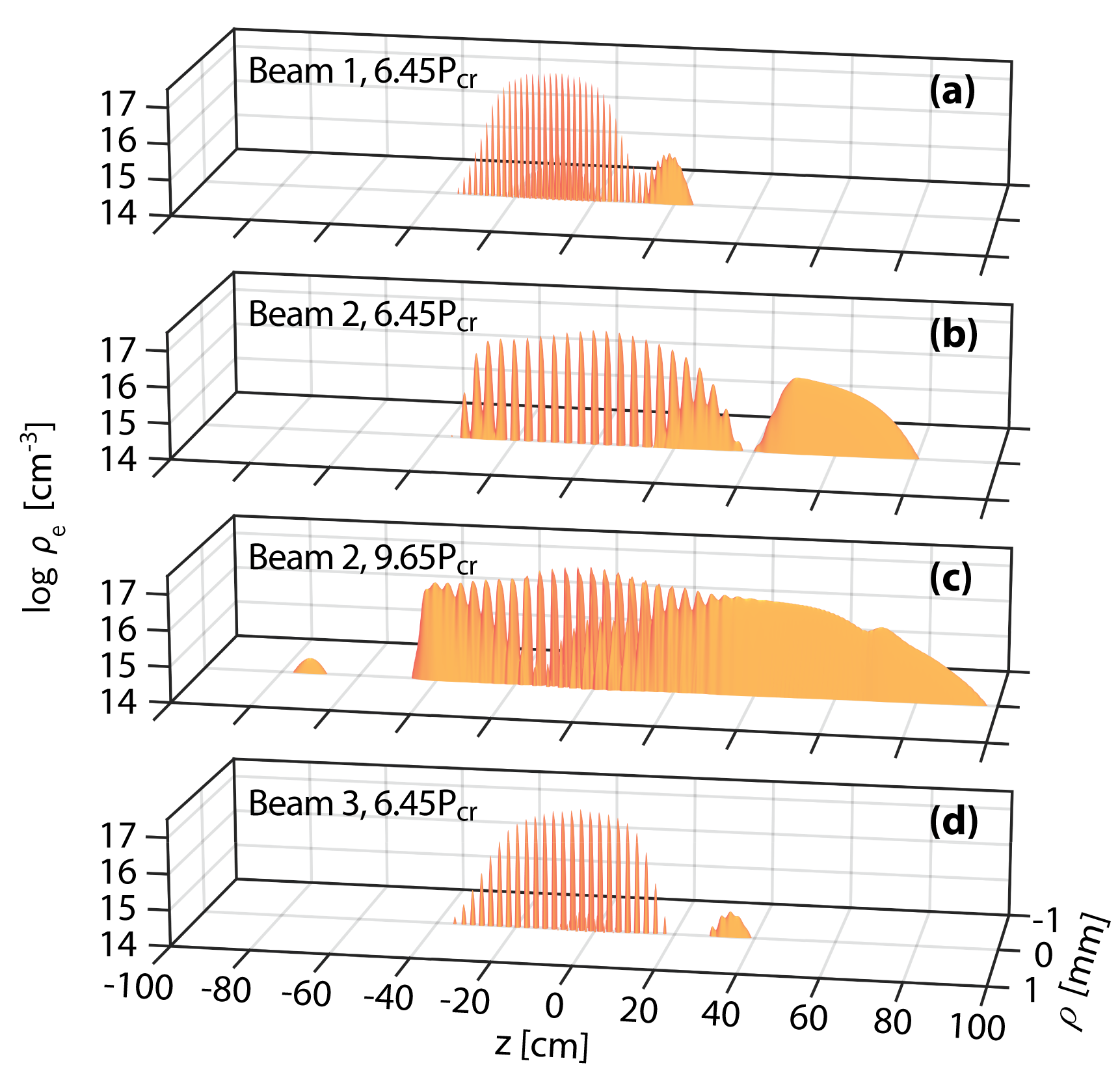}
	\caption{Evolution of the time-integrated plasma density for (a) \emph{Beam~1} with an input power of $6.45\times P_{\text{cr}}$, (b) \emph{Beam~2} with an input power of $6.45\times P_{\text{cr}}$, (c) \emph{Beam~2} with an input power of $9.65\times P_{\text{cr}}$, and (d) \emph{Beam~3} with an input power of $6.45\times P_{\text{cr}}$. Here, $\rho$ denotes the radial distance from the propagation axis, $z$.}
	\label{fig3}
\end{figure}

The evolution of the time-integrated plasma density, over the temporal extension of the pulse, for the three beams and the two input pulse energies are shown in Fig.~\ref{fig3}. It is seen that the generated electron density during the filamentation of the beams closely follows the evolution of the intensity, and the peak electron density is similar across all cases, on the order of $10^{17}$~cm$^{-3}$. 

{Figure~\ref{fig4} illustrates the evolution of axial intensity and electron density along the propagation direction for the four cases. The figure shows the modulated filamentary propagation and the corresponding modulation of electron density, allowing for comparison between cases. To facilitate a quantitative comparison of different circumstances,  several parameters of the evolution of the axial intensity during the filamentation of the beams, such as maximum intensity, $I_\text{max}$, filamentary propagation length, $L_\text{f}$, onset position of filamentation, $z_\text{o}$, and average modulation depth, $(\Delta I)_\text{avg}$, are given in Table~\ref{tab2}. 
Here, the filamentation length is defined as a length over which the envelope of the axial intensity profile exceeds a threshold value of $1.1 \times 10^{13}$~W/cm$^{2}$. Moreover, since the modulation depth is not constant over the filamentation length, the modulation depth has to be averaged over the modulated region of the filamentary propagation range. 
\begin{figure}[t]
	\centering
	\includegraphics[width=1\linewidth]{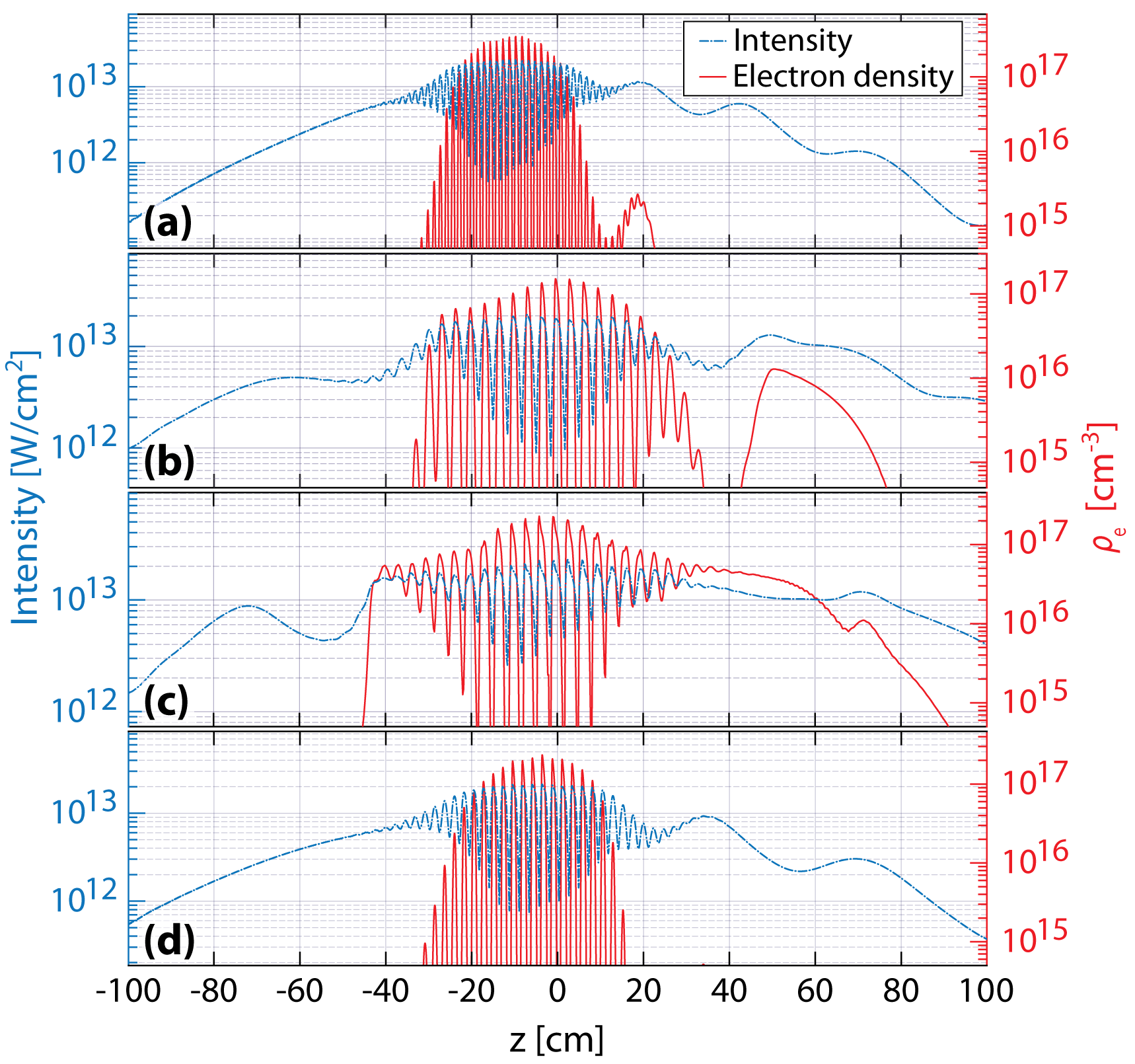}
	\caption{Axial intensity and electron density along propagation for (a) \emph{Beam~1} with an input power of $6.45\times P_{\text{cr}}$ (b) \emph{Beam~2} with an input power of $6.45\times P_{\text{cr}}$, (c) \emph{Beam~2} with an input power of $9.65\times P_{\text{cr}}$, and (d) \emph{Beam~3} with an input power of $6.45\times P_{\text{cr}}$.}
	\label{fig4}
\end{figure}

The modulation period of the plasma density consistently matches that of the corresponding intensity in all cases. As the input pulse power for \emph{Beam~2} increases, the filamentation length extends, and the onset of filamentation moves to a shorter propagation distance. Meanwhile, the average modulation depth of intensity decreases. Notably, the maximum intensity, $I_{\text{max}}$, remains clamped at approximately $2\times10^{13}$~W/cm$^2$ indicating that filamentation occurs in all scenarios.

	\begin{table}[t]
		\caption{Parameters of the axial intensity profiles illustrated in Fig.~\ref{fig4}~(a-d)}
		\label{tab2}
		\begin{ruledtabular}
		\squeezetable
		\begin{tabular}{lcllll}
		
			Beam & \begin{tabular}{c}Input Power \\ $[\times P_{\text{cr}}]$ \end{tabular}& \begin{tabular}{c}$L_\text{f}$ \\ $\text{[cm]}$ \end{tabular} & \begin{tabular}{c} $z_\text{o}$ \\ $\text{[cm]}$ \end{tabular} & \begin{tabular}{c} $I_\text{max}$ \\ $\mathrm{[W/cm^2]}$ \end{tabular} & \begin{tabular}{c} $(\Delta I)_\text{avg}$ \\ $\mathrm{[W/cm^2]}$ \end{tabular}\\
\hline
\\[-5pt]
			\emph{Beam~1}& 6.45 & 40 & -30 &2.2$\times 10^{13}$&8.2$\times 10^{12}$\\
			\emph{Beam~2}& 6.45 & 58 & -33 &2.0$\times 10^{13}$&7.0$\times 10^{12}$\\
			\emph{Beam~2}& 9.65 & 88 & -44 &2.3$\times 10^{13}$&4.8$\times 10^{12}$\\
			\emph{Beam~3}& 6.45 & 46 & -30 &2.1$\times 10^{13}$&8.0$\times 10^{12}$\\

		\end{tabular}
		\end{ruledtabular}
	\end{table}
	
		\begin{table}[ht]
		\caption{Nonlinear modulation period, maximum electron density, and average modulation depth of the axial electron density profiles illustrated in Fig.~\ref{fig4}~(a-d)}
		\label{tab3}
		\begin{ruledtabular}
			\squeezetable
			\begin{tabular}{lclll}
				Beam & \begin{tabular}{c}Input Power \\ $[\times P_{\text{cr}}]$ \end{tabular} & \begin{tabular}{c}$\Lambda_\text{NL}$ \\ $\text{[mm]}$ \end{tabular}& \begin{tabular}{c}$\rho_\text{e,max}$ \\ $\mathrm{[cm^{-3}]}$ \end{tabular} & 
				\begin{tabular}{c}$(\Delta\rho_\mathrm{e})_\mathrm{avg}$ \\ $\mathrm{[cm^{-3}]}$ \end{tabular}\\[5pt]
				\hline
				\\[-5pt]
				\emph{Beam~1}& 6.45  & 14.8 &3.5$\times 10^{17}$ & 8.7$\times 10^{16}$\\
				\emph{Beam~2}& 6.45  & 33.0 &1.5$\times 10^{17}$ & 4.6$\times 10^{16}$\\
				\emph{Beam~2}& 9.65  & 33.0 &2.3$\times 10^{17}$ & 4.7$\times 10^{16}$\\
				\emph{Beam~3}& 6.45  & 23.0 &2.3$\times 10^{17}$ & 5.9$\times 10^{16}$\\
				
			\end{tabular}
		\end{ruledtabular}
	\end{table}

Also, several parameters of the axial electron density such as the nonlinear modulation period, $\Lambda_{\text{NL}}$, maximum value of the axial electron density envelope, $\rho_{\text{e,max}}$, and average modulation depth, $(\Delta\rho_\text{e})_\text{avg}$ for the four cases are measured and summarized in Table~\ref{tab3}. Interestingly, the nonlinear modulation periods are almost identical to the linear ones. This suggests that the nonlinear propagation of the beams is strongly shaped by their linear propagation characteristics. Additionally, it is observed that all beams exhibit significant modulation depths of the axial electron density, exceeding $4.5~\times~10^{16}$~cm$^{-3}$. However, the modulation depth for the beam with larger periodicity (i.e., \emph{Beam~2}) is approximately half as large. A similar trend of lower modulation depth of the beam with larger periodicity is also evident in the axial intensity profiles. This suggests that achieving higher modulation depths is easier with shorter modulation periods, This occurs because, for beams with shorter modulation periods, the accumulated nonlinear phase due to the Kerr effect during the high-intensity half modulation cycle is relatively small and insufficient to trigger self-focusing. Consequently, the intensity\textendash and thus the electron density\textendash within the low-intensity regions remain far below their respective maximum level. Moreover, as detailed in Appendix~\ref{AppndxA}, the larger-spot Bessel component of \emph{Beam~2} (i.e., $s_{1} = 90~\mu$m) would individually undergo unsteady filamentation characterized by irregular modulations of both intensity and electron density. Since this component of \emph{Beam~2} carries a large part of the total pulse energy, and it extends beyond the interference zone of the two components, it would generate a background high intensity (and consequently high electron density) zone that further decreases the modulation depths.
 
Note that the intensity clamping at approximately $2\times10^{13}$~W/cm$^{2}$ imposes an upper limit on the maximum achievable plasma density modulation depth capping it at $10^{17}$~cm$^{-3}$ in air at atmospheric pressure. However, higher modulation depths remain possible in alternative propagation media or under higher pressures. 

We also examine how the modulation depth varies with input pulse energy for \emph{Beam 1}. The results of this investigation are presented and discussed in Appendix~\ref{AppndxB}. Our findings show that increasing the input pulse energy does not enhance the modulation depths; instead, it degrades the modulation once the pulse peak power exceeds $\sim$$13~P_{\text{cr}}$ due to early filamentation of the sub-beams, which disrupts the linear power flux towards the superposition zone required to achieve a high-contrast interference of the two constituent sub-beams. This degradation is expected to be even more pronounced for longer-period beams (e.g., \emph{Beam 2}), which have larger-spot sub-beams that are more susceptible to early Kerr self-focusing and unsteady filamentation.

\begin{figure}[t]
	\centering
	\includegraphics[width=1\linewidth]{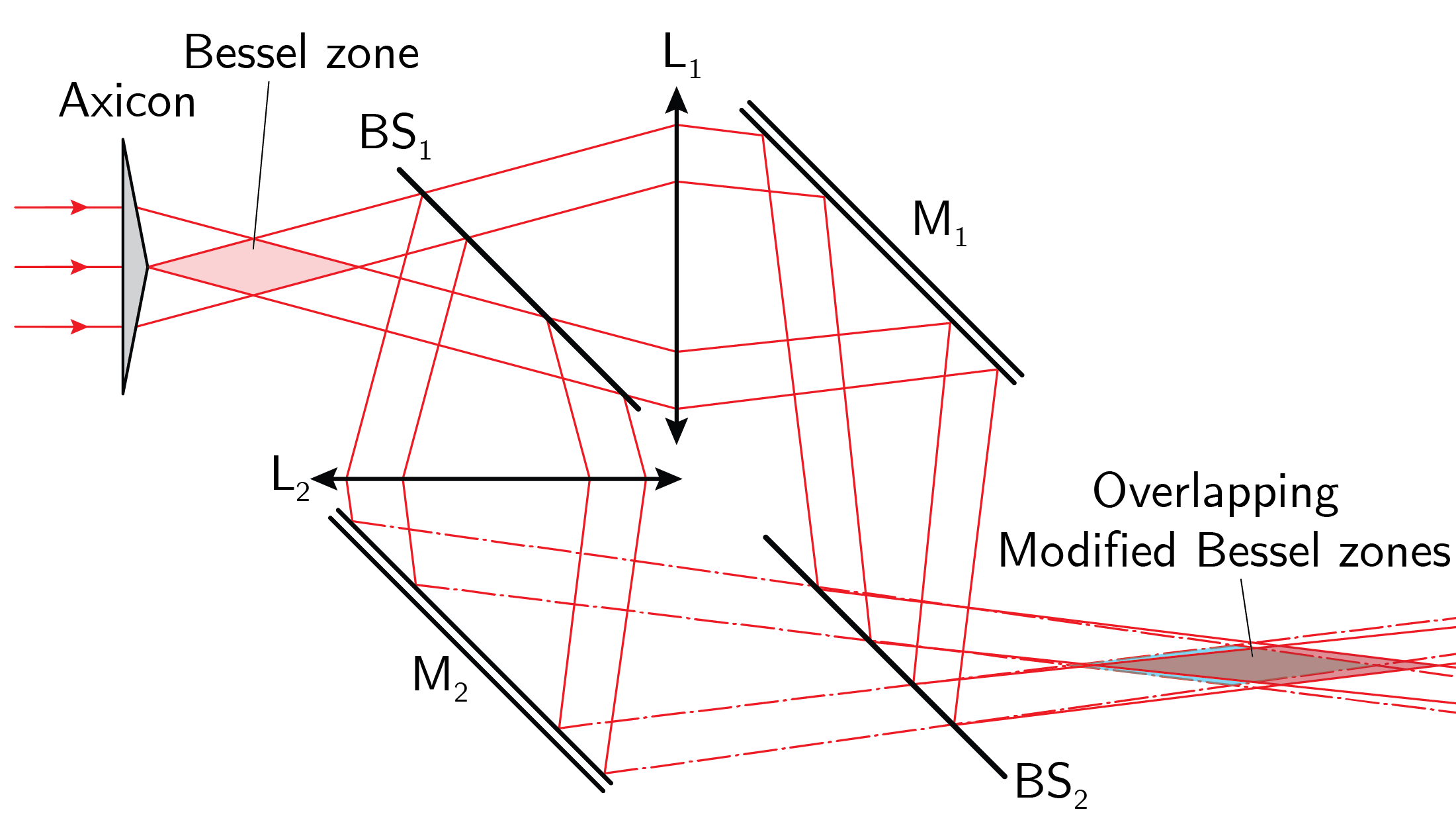}
	\caption{An interferometric configuration proposed for generating two superposed Bessel beams. An original Bessel beam is generated by an axicon. A converging lens in each arm of the interferometer (i.e., L$_{1}$, and L$_{2}$) relays the original beam to an axial distance beyond the second beam splitter, BS$_{2}$. The parameters of the relayed beams can be tuned by choosing different focal lengths for the two relay lenses or by changing the distances of the relay lenses to the first beam splitter, BS$_{1}$. Superposition of the two relayed modified Bessel beams occurs after the second beam splitter, BS$_2$. M$_{1}$, and M$_{2}$ are flat mirrors.}
	\label{fig5}
\end{figure}

From the practical point of view, two interfering Bessel beams with different central spot sizes can be generated using 
an interferometric configuration such as the one depicted in Fig.~\ref{fig5}, employing a single axicon and without a need for an SLM. In this configuration, a single original Bessel beam is generated by an axicon, and its far-field ring-shaped distribution is split by a beam splitter (BS$_1$) into the two arms of an interferometer (e.g., Mach-Zehnder type). A converging lens in each arm (L$_1$ or L$_2$), relays the Bessel beam to an axial location after recombination at a second beam splitter (BS$_2$). The parameters of the relayed Bessel beams can be tuned by adjusting the distance between the relay lens and the original Bessel zone, or by selecting appropriate focal lengths for the two relay lenses. In all cases, the two relayed Bessel beams must propagate collinearly, and their individual modified Bessel zones (i.e., depths of fields) must overlap. In the overlapping region, the two beams superpose to form an axial intensity modulation, which is essential for generating an axially modulated plasma structure.

	It is important to choose the parameters of the initial Bessel beam so as to avoid unsteady filamentation or high-density plasma generation (i.e., beyond $10^{16}$~cm$^{-3}$). Previous studies have shown that, for certain ranges of the cone angles of the initial Bessel beam and width of the incident Gaussian beam, unsteady Bessel filamentation and high-density plasma generation--which lead to spatiotemporal beam distortion--can be avoided even at relatively high input energies (on the order of a few mJ) \cite{PorrasPRL04, PolesanaPRA08, CouaironSPIE13}. To ensure that the initial Bessel beam formed by the axicon does not undergo unsteady filamentation, we investigated its nonlinear propagation. In this study, a collimated Gaussian femtosecond beam with a width of 10~mm, a pulse width of $t_0 = 50$~fs and a pulse energy of 1.53~mJ was incident on an axicon with a refractive index of 1.5, and an apex angle of 179$^\circ$. This axicon generated a Bessel beam with a cone angle of 0.25$^\circ$. We numerically simulated the nonlinear propagation of the generated beam, with the physical parameters provided in Table~\ref{tab1}, from the tip of the axicon over several meters in air. The $y-z$ cross-sectional views of the time-averaged intensity and time-integrated plasma density along the propagation are presented in Figs.~\ref{fig6}(a), and \ref{fig6}(b), respectively. The results show that the maximum intensity remains below $8\times10^{12}$~W/cm$^2$, and maximum electron density does not exceed $\sim$$9~\times~10^{13}$~cm$^{-3}$. These findings are consistent with previously reported values for the steady filamentation of Bessel beams in air \cite{PolesanaPRA08}, where it has been shown that nonlinear propagation in this regime does not lead to significant spatiotemporal evolution of the pulsed Bessel beam.
	
	\begin{figure}[t]
		\centering
		\includegraphics[width=1\linewidth]{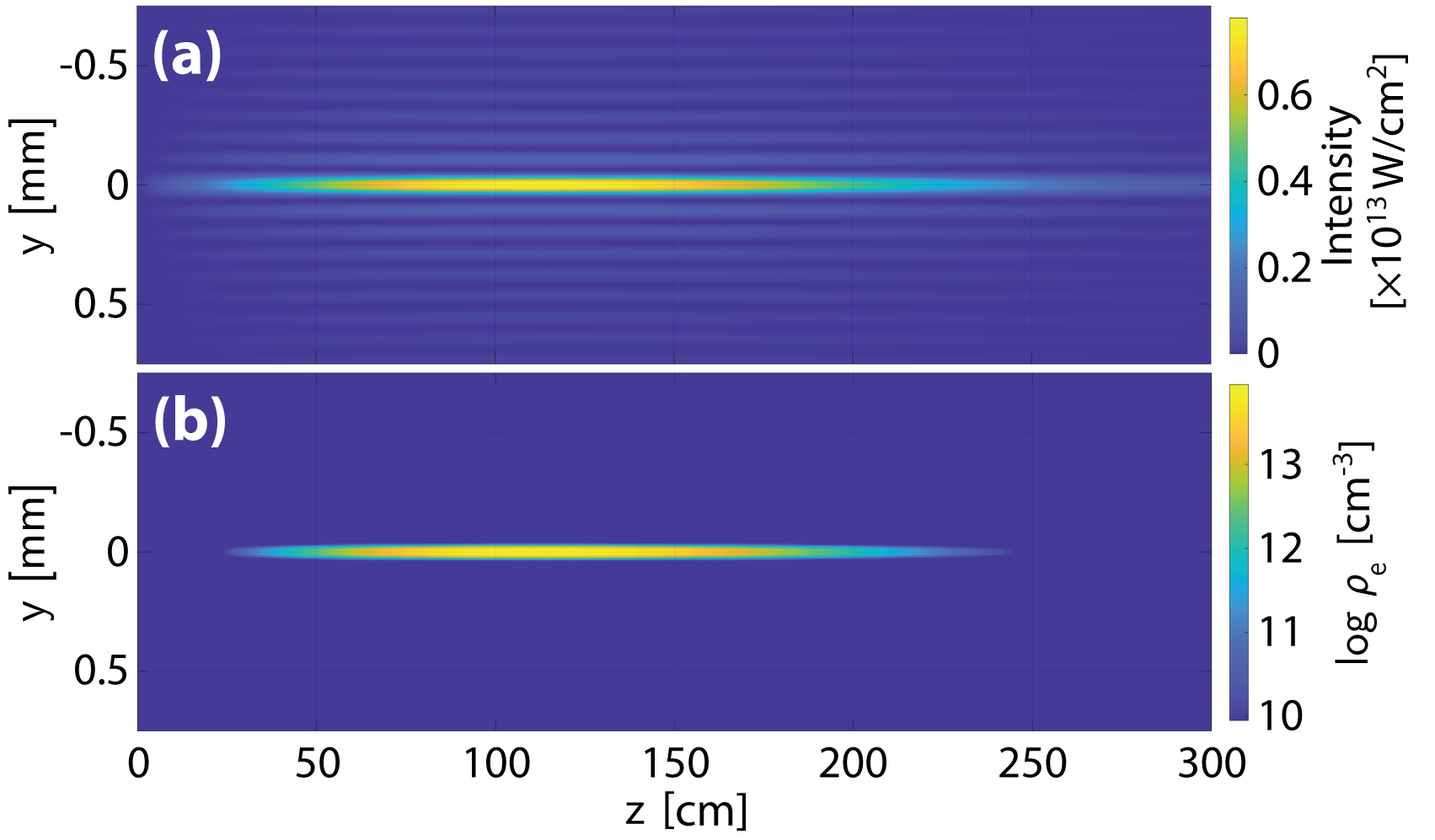}
		\caption{Cross-sectional views of the intensity (a) and plasma density (b) during the nonlinear propagation of a Bessel beam with a cone angle of 0.25$^\circ$, a pulse width of $t_0 = 50$~fs and a pulse energy of 1.53~mJ, generated after the axicon in Fig.~\ref{fig5}.}
		\label{fig6}
	\end{figure}

\section{\label{sec4} Conclusion} 
	Through numerical simulations, we demonstrated that the filamentation of two collinearly propagating femtosecond Bessel beams results in axial modulation of intensity and plasma density with significant modulation depths. We examined the characteristics of these axial modulations--including their depths and periods--for three beams and two different pulse energies. Our analysis revealed that the modulation periods for intensity and plasma density are identical across all beams and pulse energies, and these periods closely match those observed in the axial intensity modulation under linear propagation conditions. This highlights the crucial role of linear power flux in governing extreme nonlinear propagation of ultrashort interfering Bessel beams.
	
	Furthermore, we show that the modulation period can be tuned by adjusting the central spot sizes of the interfering Bessel beams or by employing appropriate external focusing. During the filamentation process, we observed intensity clamping, with peak electron densities falling within the expected range for femtosecond pulse filamentation in air.
	
	Additionally, we propose a simple interferometric setup using a single axicon to generate interfering Bessel beams, thereby producing a plasma string with high axial modulation depth. The resulting modulated plasma string, with modulation depths on the order of $10^{16}~-~10^{17}$~cm$^{-3}$ and a tunable modulation period, can be leveraged for various applications, including efficient THz-, or harmonic-generation. 

	Finally, although this study focuses on the filamentation of interfering Bessel beams in air, we expect that similar phenomena could occur in transparent solids such as glass. Such effects may enable the fabrication of periodic microstructures for photonic devices, precise volumetric machining, or improved ablation control. However, quantitative predictions would require medium-specific investigations.

\pagebreak
\onecolumngrid
\appendix
\section{Filamentation of individual constituent Bessel beams
\label{AppndxA}}
To acquire insight into the possible correlation between the filamentation of interfering Bessel beams and that of their individual constituent components, we analyzed the nonlinear propagation of the sub-beams comprising each of the three composite beams---\emph{Beam~1}, \emph{Beam~2}, and \emph{Beam~3}---at a total pulse energy of 1.03~mJ. In addition, \emph{Beam~2} was also studied at a higher energy of 1.53~mJ.  For each case, we calculated the energy content of the constituent Bessel beams and numerically investigated their filamentation using the model and method described in Section~\ref{sec2}, with physical parameters listed in Table~\ref{tab1}. An overview of the different cases and their corresponding beam parameters is provided in Table~\ref{tab4}. In all scenarios, the larger sub-beam carries $ (5-8)\times P_\mathrm{cr}$, while the smaller sub-beam carries approximately ($1-1.6)\times P_\mathrm{cr}$.
\begin{table*}[b]
	\caption{Beam parameters used in the numerical investigation of the filamentation of each sub-beam of the three composite beams and the two total pulse energies.}
	\label{tab4}
	\begin{ruledtabular}
		\begin{tabular}{lcccccc}
			Beam & \begin{tabular}{c}Total Energy \\ $\mathrm{[mJ]}$ \end{tabular} & \begin{tabular}{c} Total Power \\ $[\times P_{\text{cr}}]$ \end{tabular} & \begin{tabular}{c}Sub-beam Size\\ $[\mu$m$]$ \end{tabular} & \begin{tabular}{c}Sub-beam Cone Angle \\ $\mathrm{[deg]}$ \end{tabular} & \begin{tabular}{c}Energy \\ $\mathrm{[mJ]}$ \end{tabular} & \begin{tabular}{c}Power \\ $[\times P_{\text{cr}}]$ \end{tabular} \\
			\hline
			\multirow{2}{*}{\emph{Beam~1}} & \multirow{2}{*}{1.03} & \multirow{2}{*}{6.45} & $s_1$ = 60 & 0.12 & 0.86 & 5.38 \\
			& & & $s_2$ = 12  & 0.61 & 0.17 & 1.07 \\
			\hline
			\multirow{2}{*}{\emph{Beam~2}} & \multirow{2}{*}{1.03} & \multirow{2}{*}{6.45} & $s_1$ = 90  & 0.08 & 0.86 & 5.38 \\
			& & & $s_2$ = 18  & 0.40 & 0.17 & 1.07 \\
			\hline
			\multirow{2}{*}{\emph{Beam~2} (alt.)} & \multirow{2}{*}{1.53} & \multirow{2}{*}{9.65} & $s_1$ = 90  & 0.08 & 1.28 & 8.07 \\
			& & & $s_2$ = 18 & 0.40 & 0.25 & 1.58 \\
			\hline
			\multirow{2}{*}{\emph{Beam~3}} & \multirow{2}{*}{1.03} & \multirow{2}{*}{6.45} & $s_1$ = 75 & 0.10 & 0.86 & 5.38 \\
			& & & $s_2$ = 15  & 0.49 & 0.17 & 1.07 \\
		\end{tabular}
	\end{ruledtabular}
\end{table*} 
 
The results of these numerical simulations are shown in Fig.~\ref{fig7}. For each beam, the central cross-sectional propagation profiles of intensity and plasma density are presented in columns (a) and (b), respectively. The profiles are labeled as (a$j-i$) and (b$j-i$), where $j$ indicates the case number, and $i$ refers to the sub-beam: $i = 1$ corresponds to the sub-beam with the larger central spot size $s_1$, and $i = 2$ to the one with the smaller spot size $s_2$.
	
The presented results are in excellent quantitative and qualitative agreement with the previous studies on the filamentation of Bessel beams in air \cite{PolesanaPRA08,RoskeyOE07}, in terms of the dynamics of the nonlinear propagation, and maximum electron density and intensity (or fluence) achieved during the propagation of fs Bessel beams with comparable physical parameters. It is observed that the large sub-beam of \emph{Beam~2} (cone angle 0.08$^\circ$, $s_1 = 90~\mu$m) undergoes unsteady filamentation at both 1.03~mJ and 1.53~mJ total pulse energies (see Fig.~\ref{fig7}(a2-1) and Fig.~\ref{fig7}(a3-1)). This behavior arises because, at high input powers exceeding $5\times P_\text{cr}$,
the large-core Bessel beam undergoes Kerr self-focusing, leading to the formation of smaller, high-intensity local spots. These in turn trigger local plasma generation with densities reaching approximately $10^{16}$~cm$^{-3}$ (Fig.~\ref{fig7}(b2-1) and Fig.~\ref{fig7}(b3-1)), resulting in plasma defocusing and a modulated propagation profile.
Notably, as shown in Fig.~\ref{fig7}(a3-1), increasing the input power reduces the contrast between adjacent axial intensity peaks,
leading to the formation of an almost continuous plasma string (Fig.~\ref{fig7}(b3-1)). Below a certain input power, Kerr self-focusing may still
locally compress the beam core, but the resulting intensities remain insufficient to initiate multiphoton ionization. In this case, plasma does not form, and diffraction becomes the dominant counterbalance to self-focusing. This regime, characterized by irregular axial intensity modulations in the absence of relatively high-density plasma, has been previously observed in~\cite{RoskeyOE07} and is referred to as the \emph{weakly nonlinear propagation regime}~\cite{PolesanaPRA08}.
	
In all other cases, steady nonlinear propagation is observed. Interestingly, even for relatively large-core Bessel beams (i.e., $s_1 = 60~\mu$m and $s_1 = 75~\mu$m), the maximum intensity remains about $1\times 10^{13}$~W/cm$^2$ (see Fig.~\ref{fig7}(a1-1) and Fig.~\ref{fig7}(a4-1)), and the maximum electron density does not exceed $\sim$$1\times10^{15}$~cm$^{-3}$ (Fig.~\ref{fig7}(b1-1) and Fig.~\ref{fig7}(b4-1)). The maximum intensity and electron density for the small-core Bessel beam components of three beams at both total input pulse energies, remain below $10^{13}$~W/cm$^{2}$, and $10^{14}$~cm$^{-3}$, respectively.

A comparison of the results presented in Fig.~\ref{fig7} with those corresponding to the filamentation of superimposed Bessel beams shows that the high-frequency modulations are solely due to interference of the Bessel sub-beams in each case. Nevertheless, it can be inferred that the smooth slowly-varying axial modulations, both before and after the interference-induced rapid modulations, which are observed in the intensity and plasma density profiles of the three composite beams, especially \emph{Beam~2}, (see Fig.~\ref{fig2} and Fig.~\ref{fig3}) originate from the filamentation of higher-energy larger sub-beam, as it is longitudinally extended beyond the overlapping region of the two sub-beams and it carries higher power. 
		
\begin{figure*}[h!]
	\centering
	\includegraphics[width=0.86\linewidth]{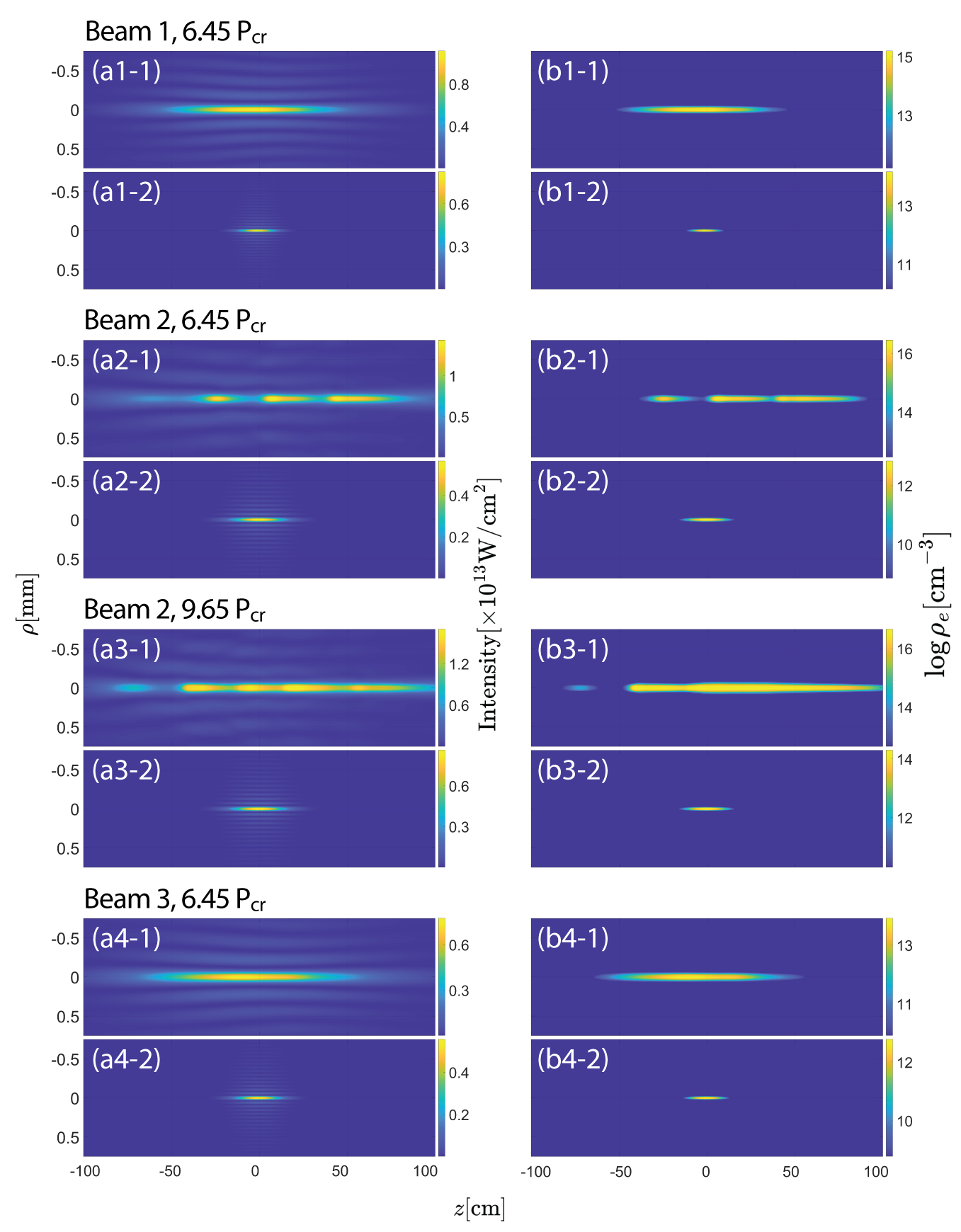}
	\caption{Nonlinear propagation of individual sub-beams of the three composite beams (i.e., \emph{Beam~1}, \emph{Beam~2}, and \emph{Beam~3}), with pulse energies of 1.03~mJ, for all three beams, and 1.53~mJ for \emph{Beam~2}. Detailed beam parameters are provided in Table~\ref{tab4}. For each case, the $\rho-z$ cross-sections of the average intensity (left column, a)  and the integrated plasma density (right column, b) are presented. For each beam, the first row corresponds to the first sub-beam with the central spot size of $s_1$, and the second row corresponds to the second sub-beam with the central spot size of $s_2$.}
	\label{fig7}
\end{figure*}

\section{Modulation depth of interfering Bessel beams at higher input pulse energies}
\label{AppndxB}
 To determine if there is an upper limit on the input pulse energy for achieving plasma modulation, we examined the influence of input pulse energy on the modulation depth of intensity and plasma density, for \emph{Beam 1}. We employed the same methodology and model, with parameters outlined in Table \ref{tab1} testing input pulse energies of 1.03~mJ, 2.06~mJ, 3.09~mJ, and 5.15~mJ, corresponding to 6.45$P_{\text{cr}}$, 12.90$P_{\text{cr}}$, 19.35$P_{\text{cr}}$, and 32.25$P_{\text{cr}}$, respectively.
 
 The axial intensity and electron density profiles along the propagation direction are illustrated in Fig. \ref{fig8} for these input pulse energies. Notably, the plasma modulation depth does not surpass $\sim$$10^{17}$ cm$^{-3}$, due to intensity clamping, which prevents the formation of higher-intensity spots, and consequently elevated electron densities. 

 For input peak pulse powers up to $\sim$$13P_{\text{cr}}$ (Fig. \ref{fig8} (a)-(b)), the plasma density modulation remains intact. However, increasing the input pulse power beyond this threshold significantly degrades the modulation, particularly in the central region of the modulated area (Fig. \ref{fig8} (c)-(d)). 
 
 This degradation can be explained by considering that modulation at various propagation distances results from the overlap of different transverse regions of the constituent sub-beams in the composite beams. As depicted schematically in Fig.~\ref{fig9}, modulation at shorter propagation distances arises from the interference of the inner portions of the conical-wavefront ring-shaped Gaussian sub-beams, whereas modulation at longer propagation distances originates from the interference of their outer parts. Given the Gaussian ring shape, both inner and outer portions of the sub-beams have lower intensities than the central parts of both sub-beams, which contribute to the modulation in the central axial region. 
 
 As the input pulse energy increases, the central part of one or both sub-beams may gain sufficient power to initiate filamentation before reaching the superposition area. This disrupts the linear power flux toward the interference region, thereby eroding the modulation in the central axial region. Meanwhile the inner and outer portions of the sub-beams may retain linear flux, preserving the axial modulation on the sides of the central modulated region (as observed in Fig.~\ref{fig8} (c)). With further increases in input pulse energy, a larger segment of  each sub-beam's central part undergoes filamentation, causing the modulated axial region to contract further from the center to the sides, as evidenced by comparing Fig.~\ref{fig8} (c) and (d). At even higher pulse energies, the modulation depth is expected to diminish substantially across the entire overlapping region of the sub-beams.
  
Moreover, as discussed in the main text, longer-period composite beams (e.g., \emph{Beam 2}) exhibit lower modulation depths than their shorter-period counterparts at identical initial pulse energies. Consequently, the reduction in modulation depth with increasing input energy is more pronounced for longer-period beams. This stems from two primary mechanisms: (1) longer-period beams consist of sub-beams with larger central spot sizes, rendering them more susceptible to early Kerr self-focusing and unsteady filamentation before the interference zone; and (2) their longer high-intensity half-cycles allow greater accumulation of nonlinear Kerr phase (at the same clamped maximum intensity due to filamentation), which triggers self-focusing that elevates intensity (and thus electron density via multiphoton ionization) in the low-intensity half-cycles, thereby diminishing overall modulation depth. As detailed in Appendix~\ref{AppndxA}, the larger-spot sub-beam in longer-period beams undergoes unsteady filamentation independently, producing a background high-intensity zone that extends beyond the interference region and further degrades modulation contrast. Both mechanisms intensify with higher input peak power, leading to more substantial modulation erosion in longer-period beams.

Crucially, although increasing the input pulse energy reduces the plasma modulation depth (for $P_{\text{in}}\gtrsim13P_{\text{cr}}$--in the case of \emph{Beam 1}), the upper limit of the modulation depth remains clamped at $\sim$$10^{17}$ cm$^{-3}$. This maximum is attainable in the nonlinear propagation regime with peak powers below a specific threshold (e.g., $P_{\text{in}}\simeq13P_{\text{cr}}$--for \emph{Beam 1}). Therefore, the input pulse energy must be optimized to maximize the plasma modulation depth.

\begin{figure}[h!]
	\centering
	\includegraphics[width=0.65\linewidth]{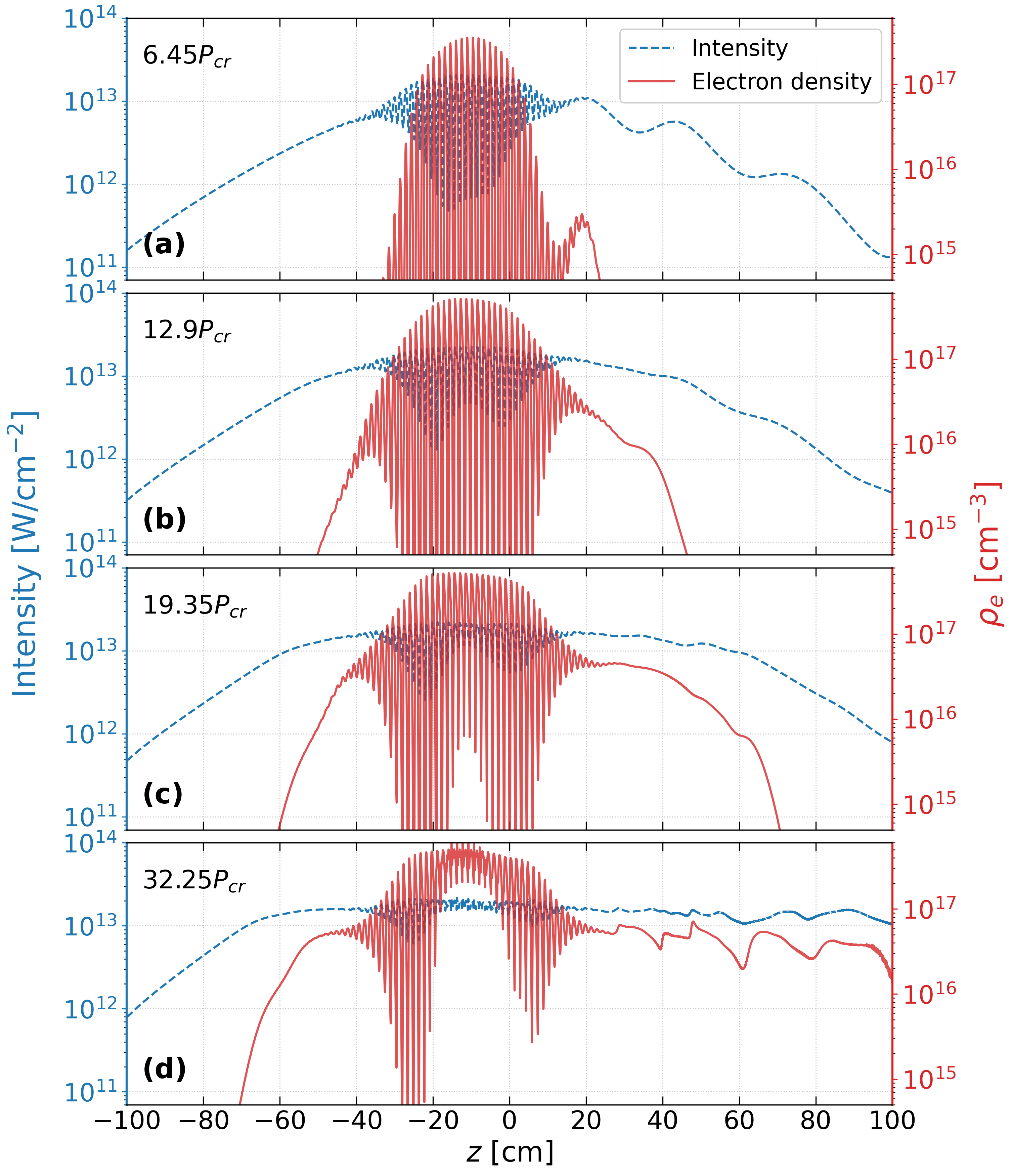}
	\caption{Axial intensity and electron density profiles for \emph{Beam 1} at input energies of 1.03~mJ, 2.06~mJ, 3.09~mJ, and 5.15~mJ ((a)-(d), respectively). The dashed blue curve represents the intensity profile (left vertical axis), while the solid red curve depicts the electron density (right vertical axis).}
	\label{fig8}
\end{figure}

\begin{figure}[h!]
	\centering
	\includegraphics[width=0.6\linewidth]{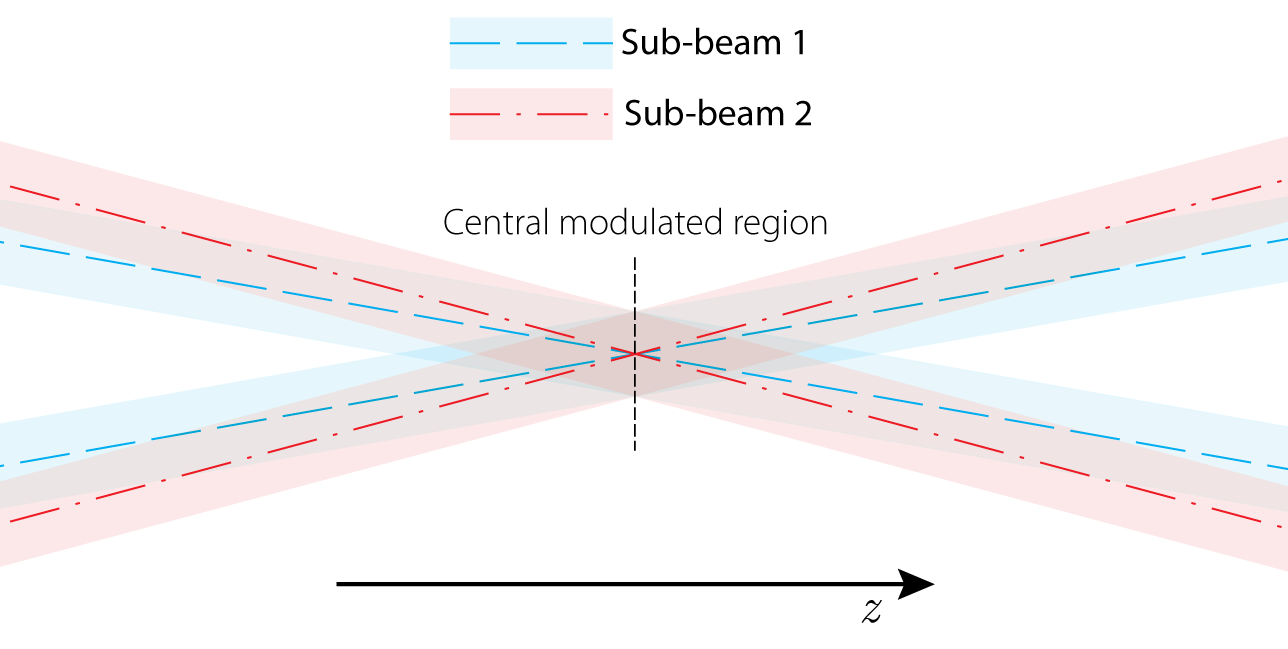}
	\caption{Schematic illustration of the superposition of the two constituent sub-beams that form the composite beams. These sub-beams are conical-wavefront Gaussian ring-shaped beams in the far-field, with different cone angles. The overlap of different sub-beam segments produces axial modulation at varying propagation distances, $z$. Specifically, the inner parts overlap at shorter distances, the central parts form the modulation in the central region, and the outer parts generate modulation at longer distances.}
	\label{fig9}
\end{figure}

\newpage
\twocolumngrid
\bibliography{References}

@article{GaizOL06,
	author = {Gaizauskas, Eugenijus and Vanagas, Egidijus and Jarutis, Vygandas and Juodkazis, Saulius and Mizeikis, Vygantas and Misawa, Hiroaki},
	date = {2006/01/01},
	doi = {10.1364/OL.31.000080},
	j2 = {Opt. Lett.},
	journal = {Optics Letters},
	journal1 = {Optics Letters},
	journal2 = {Opt. Lett.},
	journal3 = {Opt. Lett.},
	number = {1},
	pages = {80--82},
	publisher = {Optica Publishing Group},
	title = {Discrete damage traces from filamentation of Gauss-Bessel pulses},
	url = {https://opg.optica.org/ol/abstract.cfm?URI=ol-31-1-80},
	volume = {31},
	year = {2006}}

@article{AkturkOptComm09,
	author = {Akturk, Selcuk and Zhou, Bing and Franco, Michel and Couairon, Arnaud and Mysyrowicz, Andre},
	doi = {https://doi.org/10.1016/j.optcom.2008.09.048},
	issn = {0030-4018},
	journal = {Optics Communications},
	number = {1},
	pages = {129-134},
	title = {Generation of long plasma channels in air by focusing ultrashort laser pulses with an axicon},
	url = {https://www.sciencedirect.com/science/article/pii/S0030401808009607},
	volume = {282},
	year = {2009}}

@article{PorrasPRL04,
	author = {Porras, Miguel A. and Parola, Alberto and Faccio, Daniele and Dubietis, Audrius and Trapani, Paolo Di},
	doi = {10.1103/PhysRevLett.93.153902},
	issue = {15},
	journal = {Phys. Rev. Lett.},
	month = {Oct},
	numpages = {4},
	pages = {153902},
	publisher = {American Physical Society},
	title = {Nonlinear Unbalanced Bessel Beams: Stationary Conical Waves Supported by Nonlinear Losses},
	url = {https://link.aps.org/doi/10.1103/PhysRevLett.93.153902},
	volume = {93},
	year = {2004}}

@article{PolynkinOE09,
	author = {Polynkin, Pavel and Kolesik, Miroslav and Moloney, Jerome},
	doi = {10.1364/OE.17.000575},
	journal = {Opt. Express},
	month = {Jan},
	number = {2},
	pages = {575--584},
	publisher = {Optica Publishing Group},
	title = {Extended filamentation with temporally chirped femtosecond Bessel-Gauss beams in air},
	url = {https://opg.optica.org/oe/abstract.cfm?URI=oe-17-2-575},
	volume = {17},
	year = {2009}}

@article{DotaPRA12,
	author = {Dota, Krithika and Pathak, Abhishek and Dharmadhikari, J. A. and Mathur, D. and Dharmadhikari, A. K.},
	doi = {10.1103/PhysRevA.86.023808},
	issue = {2},
	journal = {Phys. Rev. A},
	month = {Aug},
	numpages = {8},
	pages = {023808},
	publisher = {American Physical Society},
	title = {Femtosecond laser filamentation in condensed media with {B}essel beams},
	url = {https://link.aps.org/doi/10.1103/PhysRevA.86.023808},
	volume = {86},
	year = {2012}}

@inproceedings{CouaironSPIE13,
	author = {Couairon, A. and Lotti, A. and Panagiotopoulos, P. and Abdollahpour, D. and Faccio, D. and Papazoglou, D. G. and Tzortzakis, S. and Courvoisier, F. and Dudley, J. M.},
	booktitle = {17th International School on Quantum Electronics: Laser Physics and Applications},
	doi = {10.1117/12.2014198},
	editor = {Dreischuh, Tanja N. and Daskalova, Albena T.},
	organization = {International Society for Optics and Photonics},
	pages = {87701E},
	publisher = {SPIE},
	title = {{Ultrashort laser pulse filamentation with Airy and Bessel beams}},
	url = {https://doi.org/10.1117/12.2014198},
	volume = {8770},
	year = {2013}}

@article{PolesanaPRA08,
	author = {Polesana, P. and Franco, M. and Couairon, A. and Faccio, D. and Di Trapani, P.},
	doi = {10.1103/PhysRevA.77.043814},
	issue = {4},
	journal = {Phys. Rev. A},
	month = {Apr},
	numpages = {11},
	pages = {043814},
	publisher = {American Physical Society},
	title = {Filamentation in Kerr media from pulsed {B}essel beams},
	url = {https://link.aps.org/doi/10.1103/PhysRevA.77.043814},
	volume = {77},
	year = {2008}}

@article{McGloinOL03,
	author = {McGloin, D. and Garc\'{e}s-Ch\'{a}vez, V. and Dholakia, K.},
	doi = {10.1364/OL.28.000657},
	journal = {Opt. Lett.},
	month = {Apr},
	number = {8},
	pages = {657--659},
	publisher = {Optica Publishing Group},
	title = {Interfering {B}essel beams for optical micromanipulation},
	url = {https://opg.optica.org/ol/abstract.cfm?URI=ol-28-8-657},
	volume = {28},
	year = {2003}}

@article{Durnin87,
	author = {Durnin, J.},
	doi = {10.1364/JOSAA.4.000651},
	journal = {J. Opt. Soc. Am. A},
	month = {Apr},
	number = {4},
	pages = {651--654},
	publisher = {Optica Publishing Group},
	title = {Exact solutions for nondiffracting beams. {I}. The scalar theory},
	url = {https://opg.optica.org/josaa/abstract.cfm?URI=josaa-4-4-651},
	volume = {4},
	year = {1987}}

@article{PearsonPRE11,
	author = {Pearson, Andrew J. and Palastro, John and Antonsen, Thomas M.},
	date = {2011/05/25/},
	day = {25},
	doi = {10.1103/PhysRevE.83.056403},
	id = {10.1103/PhysRevE.83.056403},
	j1 = {PRE},
	journal = {Physical Review E},
	journal1 = {Phys. Rev. E},
	month = {05},
	number = {5},
	pages = {056403--},
	publisher = {American Physical Society},
	title = {Simulation of terahertz generation in corrugated plasma waveguides},
	volume = {83},
	year = {2011}}

@article{DahiyaAPL10,
	author = {Dahiya, Deepak and Sajal, Vivek and Sharma, A. K.},
	doi = {10.1063/1.3291674},
	issn = {0003-6951},
	journal = {Applied Physics Letters},
	month = {01},
	number = {2},
	pages = {021501},
	title = {{Self-injection of electrons in a laser-wakefield accelerator by using longitudinal density ripple}},
	volume = {96},
	year = {2010}}

@article{KatTNS85,
	author = {Katsouleas, T. and Dawson, J. M. and Sultana, D. and Yan, Y. T.},
	doi = {10.1109/TNS.1985.4334426},
	isbn = {1558-1578},
	journal = {IEEE Transactions on Nuclear Science},
	journal1 = {IEEE Transactions on Nuclear Science},
	journal2 = {IEEE Transactions on Nuclear Science},
	number = {5},
	pages = {3554--3556},
	title = {A Side-Injected-Laser Plasma Accelerator},
	vo = {32},
	volume = {32},
	year = {1985},
	year1 = {Oct. 1985}}

@article{YoonPRL14,
	author = {Yoon, S. J. and Palastro, J. P. and Milchberg, H. M.},
	date = {2014/04/03/},
	day = {03},
	doi = {10.1103/PhysRevLett.112.134803},
	id = {10.1103/PhysRevLett.112.134803},
	j1 = {PRL},
	journal = {Physical Review Letters},
	journal1 = {Phys. Rev. Lett.},
	month = {04},
	number = {13},
	pages = {134803--},
	publisher = {American Physical Society},
	title = {Quasi-Phase-Matched Laser Wakefield Acceleration},
	volume = {112},
	year = {2014}}

@article{LayerPRL07,
	author = {Layer, B. D. and York, A. and Antonsen, T. M. and Varma, S. and Chen, Y. -H. and Leng, Y. and Milchberg, H. M.},
	date = {2007/07/19/},
	day = {19},
	doi = {10.1103/PhysRevLett.99.035001},
	id = {10.1103/PhysRevLett.99.035001},
	j1 = {PRL},
	journal = {Physical Review Letters},
	journal1 = {Phys. Rev. Lett.},
	month = {07},
	number = {3},
	pages = {035001--},
	publisher = {American Physical Society},
	title = {Ultrahigh-Intensity Optical Slow-Wave Structure},
	volume = {99},
	year = {2007}}

@article{ShiPRL11,
	author = {Shi, Liping and Li, Wenxue and Wang, Yongdong and Lu, Xin and Ding, Liang'en and Zeng, Heping},
	date = {2011/08/23/},
	day = {23},
	doi = {10.1103/PhysRevLett.107.095004},
	id = {10.1103/PhysRevLett.107.095004},
	j1 = {PRL},
	journal = {Physical Review Letters},
	journal1 = {Phys. Rev. Lett.},
	month = {08},
	number = {9},
	pages = {095004--},
	publisher = {American Physical Society},
	title = {Generation of High-Density Electrons Based on Plasma Grating Induced Bragg Diffraction in Air},
	volume = {107},
	year = {2011}}

@article{LehmannPRE19,
	author = {Lehmann, G. and Spatschek, K. H.},
	date = {2019/09/06/},
	day = {06},
	doi = {10.1103/PhysRevE.100.033205},
	id = {10.1103/PhysRevE.100.033205},
	j1 = {PRE},
	journal = {Physical Review E},
	journal1 = {Phys. Rev. E},
	month = {09},
	number = {3},
	pages = {033205--},
	publisher = {American Physical Society},
	title = {Plasma volume holograms for focusing and mode conversion of ultraintense laser pulses},
	volume = {100},
	year = {2019}}

@article{EdwardsPRAp22,
	author = {Edwards, Matthew R. and Michel, Pierre},
	date = {2022/08/09/},
	day = {09},
	doi = {10.1103/PhysRevApplied.18.024026},
	id = {10.1103/PhysRevApplied.18.024026},
	j1 = {PRAPPLIED},
	journal = {Physical Review Applied},
	journal1 = {Phys. Rev. Appl.},
	month = {08},
	number = {2},
	pages = {024026--},
	publisher = {American Physical Society},
	title = {Plasma Transmission Gratings for Compression of High-Intensity Laser Pulses},
	volume = {18},
	year = {2022}}

@article{SuntsovAPL09,
	author = {Suntsov, S. and Abdollahpour, D. and Papazoglou, D. G. and Tzortzakis, S.},
	doi = {10.1063/1.3157908},
	isbn = {0003-6951},
	journal = {Applied Physics Letters},
	journal1 = {Appl. Phys. Lett.},
	month = {11/11/2023},
	number = {25},
	pages = {251104},
	title = {Femtosecond laser induced plasma diffraction gratings in air as photonic devices for high intensity laser applications},
	volume = {94},
	year = {2009}}

@article{SuntsovPRA10,
	author = {Suntsov, S. and Abdollahpour, D. and Papazoglou, D. G. and Tzortzakis, S.},
	doi = {10.1103/PhysRevA.81.033817},
	issue = {3},
	journal = {Phys. Rev. A},
	month = {Mar},
	numpages = {4},
	pages = {033817},
	publisher = {American Physical Society},
	title = {Filamentation-induced third-harmonic generation in air via plasma-enhanced third-order susceptibility},
	url = {https://link.aps.org/doi/10.1103/PhysRevA.81.033817},
	volume = {81},
	year = {2010}}

@article{SuntsovOE09,
	author = {Suntsov, S. and Abdollahpour, D. and Papazoglou, D. G. and Tzortzakis, S.},
	date = {2009/03/02},
	doi = {10.1364/OE.17.003190},
	j2 = {Opt. Express},
	journal = {Optics Express},
	journal1 = {Optics Express},
	journal2 = {Opt. Express},
	journal3 = {Opt. Express},
	number = {5},
	pages = {3190--3195},
	publisher = {Optica Publishing Group},
	title = {Efficient third-harmonic generation through tailored IR femtosecond laser pulse filamentation in air},
	url = {https://opg.optica.org/oe/abstract.cfm?URI=oe-17-5-3190},
	volume = {17},
	year = {2009}}

@article{SteinNJP11,
	author = {Steingrube, D S and Schulz, E and Binhammer, T and Gaarde, M B and Couairon, A and Morgner, U and Kova{\v c}ev, M},
	doi = {10.1088/1367-2630/13/4/043022},
	journal = {New Journal of Physics},
	month = {apr},
	number = {4},
	pages = {043022},
	title = {High-order harmonic generation directly from a filament},
	url = {https://dx.doi.org/10.1088/1367-2630/13/4/043022},
	volume = {13},
	year = {2011}}

@article{KasparianSci03,
	author = {Kasparian, J. and Rodriguez, M. and M{\'e}jean, G. and Yu, J. and Salmon, E. and Wille, H. and Bourayou, R. and Frey, S. and Andr{\'e}, Y.-B. and Mysyrowicz, A. and Sauerbrey, R. and Wolf, J.-P. and W{\"o}ste, L.},
	doi = {10.1126/science.1085020},
	journal = {Science},
	number = {5629},
	pages = {61-64},
	title = {White-Light Filaments for Atmospheric Analysis},
	volume = {301},
	year = {2003}}

@article{TzortzakisPRL01,
	author = {Tzortzakis, S. and Berg\'e, L. and Couairon, A. and Franco, M. and Prade, B. and Mysyrowicz, A.},
	doi = {10.1103/PhysRevLett.86.5470},
	issue = {24},
	journal = {Phys. Rev. Lett.},
	month = {Jun},
	numpages = {0},
	pages = {5470--5473},
	publisher = {American Physical Society},
	title = {Breakup and Fusion of Self-Guided Femtosecond Light Pulses in Air},
	url = {https://link.aps.org/doi/10.1103/PhysRevLett.86.5470},
	volume = {86},
	year = {2001}}

@article{MitroOptica16,
	author = {Mitrofanov, A. V. and Voronin, A. A. and Sidorov-Biryukov, D. A. and Mitryukovsky, S. I. and Fedotov, A. B. and Serebryannikov, E. E. and Meshchankin, D. V. and Shumakova, V. and Ali{\v s}auskas, S. and Pug{\v z}lys, A. and Panchenko, V. Ya. and Baltu{\v s}ka, A. and Zheltikov, A. M.},
	date = {2016/03/20},
	doi = {10.1364/OPTICA.3.000299},
	j2 = {Optica},
	journal = {Optica},
	journal1 = {Optica},
	journal2 = {Optica},
	journal3 = {Optica},
	number = {3},
	pages = {299--302},
	publisher = {Optica Publishing Group},
	title = {Subterawatt few-cycle mid-infrared pulses from a single filament},
	url = {https://opg.optica.org/optica/abstract.cfm?URI=optica-3-3-299},
	volume = {3},
	year = {2016}}

@article{KimNP08,
	author = {Kim, K. Y. and Taylor, A. J. and Glownia, J. H. and Rodriguez, G.},
	date = {2008/10/01},
	doi = {10.1038/nphoton.2008.153},
	id = {Kim2008},
	isbn = {1749-4893},
	journal = {Nature Photonics},
	number = {10},
	pages = {605--609},
	title = {Coherent control of terahertz supercontinuum generation in ultrafast laser--gas interactions},
	url = {https://doi.org/10.1038/nphoton.2008.153},
	volume = {2},
	year = {2008}}

@article{XiePRL06,
	author = {Xie, Xu and Dai, Jianming and Zhang, X.-C.},
	doi = {10.1103/PhysRevLett.96.075005},
	issue = {7},
	journal = {Phys. Rev. Lett.},
	month = {Feb},
	numpages = {4},
	pages = {075005},
	publisher = {American Physical Society},
	title = {Coherent Control of THz Wave Generation in Ambient Air},
	url = {https://link.aps.org/doi/10.1103/PhysRevLett.96.075005},
	volume = {96},
	year = {2006}}

@article{HouardNP23,
	author = {Houard, Aur{\'e}lien and Walch, Pierre and Produit, Thomas and Moreno, Victor and Mahieu, Benoit and Sunjerga, Antonio and Herkommer, Clemens and Mostajabi, Amirhossein and Andral, Ugo and Andr{\'e}, Yves-Bernard and Lozano, Magali and Bizet, Laurent and Schroeder, Malte C. and Schimmel, Guillaume and Moret, Michel and Stanley, Mark and Rison, W. A. and Maurice, Oliver and Esmiller, Bruno and Michel, Knut and Haas, Walter and Metzger, Thomas and Rubinstein, Marcos and Rachidi, Farhad and Cooray, Vernon and Mysyrowicz, Andr{\'e} and Kasparian, J{\'e}r{\^o}me and Wolf, Jean-Pierre},
	date = {2023/03/01},
	doi = {10.1038/s41566-022-01139-z},
	id = {Houard2023},
	isbn = {1749-4893},
	journal = {Nature Photonics},
	number = {3},
	pages = {231--235},
	title = {Laser-guided lightning},
	url = {https://doi.org/10.1038/s41566-022-01139-z},
	volume = {17},
	year = {2023}}

@article{GuptaPRE11,
	author = {Gupta, Devki Nandan and Nam, In Hyuk and Suk, Hyyong},
	date = {2011/11/04/},
	day = {04},
	doi = {10.1103/PhysRevE.84.056403},
	id = {10.1103/PhysRevE.84.056403},
	j1 = {PRE},
	journal = {Physical Review E},
	journal1 = {Phys. Rev. E},
	month = {11},
	number = {5},
	pages = {056403--},
	publisher = {American Physical Society},
	title = {Laser-driven plasma beat-wave propagation in a density-modulated plasma},
	ty = {JOUR},
	url = {https://link.aps.org/doi/10.1103/PhysRevE.84.056403},
	volume = {84},
	year = {2011}}

@article{LinPP06,
	author = {Lin, M. W. and Chen, Y. M. and Pai, C. H. and Kuo, C. C. and Lee, K. H. and Wang, J. and Chen, S. Y. and Lin, J. Y.},
	journal = {Physics of Plasmas},
	number = {11},
	pages = {110701-4},
	title = {Programmable fabrication of spatial structures in a gas jet by laser machining with a spatial light modulator},
	type = {Journal Article},
	url = {http://link.aip.org/link/?PHP/13/110701/1},
	volume = {13},
	year = {2006}}

@article{HineOL16,
	author = {Hine, G. A. and Goers, A. J. and Feder, L. and Elle, J. A. and Yoon, S. J. and Milchberg, H. M.},
	doi = {10.1364/OL.41.003427},
	journal = {Optics Letters},
	number = {15},
	pages = {3427-3430},
	title = {Generation of axially modulated plasma waveguides using a spatial light modulator},
	type = {Journal Article},
	url = {http://ol.osa.org/abstract.cfm?URI=ol-41-15-3427},
	volume = {41},
	year = {2016}}

@article{LayerOE09,
	author = {Layer, B. D. and York, A. G. and Varma, S. and Chen, Y. H. and Milchberg, H. M.},
	journal = {Opt. Express},
	number = {6},
	pages = {4263-4267},
	title = {Periodic index-modulated plasma waveguide},
	type = {Journal Article},
	url = {http://www.opticsexpress.org/abstract.cfm?URI=oe-17-6-4263},
	volume = {17},
	year = {2009}}

@article{AntonsenPP07,
	author = {Antonsen, Jr Thomas M. and Palastro, John and Milchberg, Howard M.},
	doi = {10.1063/1.2715864},
	journal = {Physics of Plasmas},
	number = {3},
	pages = {033107-9},
	title = {Excitation of terahertz radiation by laser pulses in nonuniform plasma channels},
	type = {Journal Article},
	url = {http://link.aip.org/link/?PHP/14/033107/1},
	volume = {14},
	year = {2007}}

@article{YorkPRL08,
	author = {York, A. G. and Milchberg, H. M. and Palastro, J. P. and Antonsen, T. M.},
	journal = {Physical Review Letters},
	number = {19},
	pages = {195001},
	title = {Direct Acceleration of Electrons in a Corrugated Plasma Waveguide},
	type = {Journal Article},
	url = {https://link.aps.org/doi/10.1103/PhysRevLett.100.195001},
	volume = {100},
	year = {2008}}

@article{CouaironOL05,
	author = {Couairon, A. and Franco, M. and Mysyrowicz, A. and Biegert, J. and Keller, U.},
	journal = {Optics Letters},
	number = {19},
	pages = {2657-2659},
	title = {Pulse self-compression to the single-cycle limit by filamentation in a gas with a pressure gradient},
	type = {Journal Article},
	url = {http://ol.osa.org/abstract.cfm?URI=ol-30-19-2657},
	volume = {30},
	year = {2005}}

@article{CouaironPhysRep07,
	author = {Couairon, A. and Mysyrowicz, A.},
	issn = {0370-1573},
	journal = {Physics Reports},
	number = {2-4},
	pages = {47-189},
	title = {Femtosecond filamentation in transparent media},
	type = {Journal Article},
	url = {http://www.sciencedirect.com/science/article/B6TVP-4N0HJBX-1/2/94b551a9e5a24405be3cd4b0758bb9ec},
	volume = {441},
	year = {2007}}

@article{CerdaOE98,
	author = {Ch\'{a}vez-Cerda, S. and Tepichin, E. and Meneses-Nava, M. A. and Ramirez, G. and Hickmann, J. Miguel},
	doi = {10.1364/OE.3.000524},
	journal = {Opt. Express},
	month = {Dec},
	number = {13},
	pages = {524--529},
	publisher = {OSA},
	title = {Experimental observation of interfering {B}essel beams},
	url = {http://www.osapublishing.org/oe/abstract.cfm?URI=oe-3-13-524},
	volume = {3},
	year = {1998}}

@article{CerdaOL98,
	author = {Ch\'{a}vez-Cerda, S. and Meneses-Nava, M. A. and Hickmann, J. Miguel},
	doi = {10.1364/OL.23.001871},
	journal = {Opt. Lett.},
	month = {Dec},
	number = {24},
	pages = {1871--1873},
	publisher = {OSA},
	title = {Interference of traveling nondiffracting beams},
	url = {http://www.osapublishing.org/ol/abstract.cfm?URI=ol-23-24-1871},
	volume = {23},
	year = {1998}}

@article{CouaironEPJ2011,
	author = {Couairon, A. and Brambilla, E. and Corti, T. and Majus, D. and de J. Ram{\'\i}rez-G{\'o}ngora, O. and Kolesik, M.},
	doi = {10.1140/epjst/e2011-01503-3},
	issn = {1951-6401},
	journal = {The European Physical Journal Special Topics},
	number = {1},
	pages = {5-76},
	title = {Practitioner's guide to laser pulse propagation models and simulation},
	type = {Journal Article},
	url = {https://doi.org/10.1140/epjst/e2011-01503-3},
	volume = {199},
	year = {2011}}

@article{CouaironPRA2003,
	author = {Couairon, A.},
	doi = {10.1103/PhysRevA.68.015801},
	issue = {1},
	journal = {Phys. Rev. A},
	month = {Jul},
	numpages = {4},
	pages = {015801},
	publisher = {American Physical Society},
	title = {Dynamics of femtosecond filamentation from saturation of self-focusing laser pulses},
	url = {https://link.aps.org/doi/10.1103/PhysRevA.68.015801},
	volume = {68},
	year = {2003}}

@article{AbdollahpourOE09,
	author = {Abdollahpour, D. and Panagiotopoulos, P. and Turconi, M. and Jedrkiewicz, O. and Faccio, D. and Trapani, P. Di and Couairon, A. and Papazoglou, D. G. and Tzortzakis, S.},
	doi = {10.1364/OE.17.005052},
	journal = {Opt. Express},
	month = {Mar},
	number = {7},
	pages = {5052--5057},
	publisher = {Optica Publishing Group},
	title = {Long spatio-temporally stationary filaments in air using short pulse {UV} laser {B}essel beams},
	url = {https://opg.optica.org/oe/abstract.cfm?URI=oe-17-7-5052},
	volume = {17},
	year = {2009}}

@article{AbdollahpourPRL10,
	author = {Abdollahpour, Daryoush and Suntsov, Sergiy and Papazoglou, Dimitrios G. and Tzortzakis, Stelios},
	doi = {10.1103/PhysRevLett.105.253901},
	issue = {25},
	journal = {Phys. Rev. Lett.},
	month = {Dec},
	numpages = {4},
	pages = {253901},
	publisher = {American Physical Society},
	title = {Spatiotemporal {A}iry Light Bullets in the Linear and Nonlinear Regimes},
	url = {https://link.aps.org/doi/10.1103/PhysRevLett.105.253901},
	volume = {105},
	year = {2010}}

@article{ZouPRA23,
	author = {Zou, Long and Sun, Chen and Yu, Jin and Couairon, Arnaud},
	doi = {10.1103/PhysRevA.108.023524},
	issue = {2},
	journal = {Phys. Rev. A},
	month = {Aug},
	numpages = {8},
	pages = {023524},
	publisher = {American Physical Society},
	title = {Laser energy deposition with ring-{A}iry beams beyond kilometer range in the atmosphere},
	url = {https://link.aps.org/doi/10.1103/PhysRevA.108.023524},
	volume = {108},
	year = {2023}}

@article{Dergachev2014,
	author = {Dergachev, A.A. and Ionin, A.A. and Kandidov, V.P. and Mokrousova, D.V. and Seleznev, L.V. and Sinitsyn, D.V. and Sunchugasheva, E.S. and Shlenov, S.A. and Shustikova, A.P.},
	doi = {10.1070/QE2014v044n12ABEH015472},
	journal = {Quantum Electronics},
	month = {dec},
	number = {12},
	pages = {1085},
	publisher = {Turpion Ltd and the Russian Academy of Sciences},
	title = {Plasma channels during filamentation of a femtosecond laser pulse with wavefront astigmatism in air},
	url = {https://dx.doi.org/10.1070/QE2014v044n12ABEH015472},
	volume = {44},
	year = {2014}}

@article{Vuong2006,
	author = {Vuong, Luat T. and Grow, Taylor D. and Ishaaya, Amiel and Gaeta, Alexander L. and 't Hooft, Gert W. and Eliel, Eric R. and Fibich, Gadi},
	doi = {10.1103/PhysRevLett.96.133901},
	issue = {13},
	journal = {Phys. Rev. Lett.},
	month = {Apr},
	numpages = {4},
	pages = {133901},
	publisher = {American Physical Society},
	title = {Collapse of Optical Vortices},
	url = {https://link.aps.org/doi/10.1103/PhysRevLett.96.133901},
	volume = {96},
	year = {2006}}

@article{EdwardsPRL24,
	author = {Edwards, M. R. and Fasano, N. M. and Giakas, A. M. and Wang, M. M. and Griff-McMahon, J. and Morozov, A. and Perez-Ramirez, V. M. and Lemos, N. and Michel, P. and Mikhailova, J. M.},
	doi = {10.1103/PhysRevLett.133.155101},
	issue = {15},
	journal = {Phys. Rev. Lett.},
	month = {Oct},
	numpages = {6},
	pages = {155101},
	publisher = {American Physical Society},
	title = {Greater than Five-Order-of-Magnitude Postcompression Temporal Contrast Improvement with an Ionization Plasma Grating},
	url = {https://link.aps.org/doi/10.1103/PhysRevLett.133.155101},
	volume = {133},
	year = {2024}}

@article{RoskeyOE07,
	author = {Roskey, D.E. and Kolesik, M. and Moloney, J.V. and Wright, E.M.},
	doi = {10.1364/OE.15.009893},
	journal = {Opt. Express},
	month = {Aug},
	number = {16},
	pages = {9893--9907},
	publisher = {Optica Publishing Group},
	title = {Self-action and regularized self-guiding of pulsed Bessel-like beams in air},
	url = {https://opg.optica.org/oe/abstract.cfm?URI=oe-15-16-9893},
	volume = {15},
	year = {2007}}

@Article{PorrasPRA15,
  author    = {Porras, Miguel A. and Ruiz-Jiménez, Carlos and Losada, Juan Carlos},
  journal   = {Phys. Rev. A},
  title     = {Underlying conservation and stability laws in nonlinear propagation of axicon-generated Bessel beams},
  year      = {2015},
  month     = dec,
  number    = {6},
  pages     = {063826},
  volume    = {92},
  doi       = {10.1103/PhysRevA.92.063826},
  publisher = {American Physical Society},
  refid     = {10.1103/PhysRevA.92.063826},
  url       = {https://link.aps.org/doi/10.1103/PhysRevA.92.063826},
}
		
\end{document}